\documentclass[12pt]{article}
\usepackage[utf8]{inputenc}
\usepackage{graphicx}
\usepackage{amsmath}
\usepackage{amssymb}
\usepackage{amsfonts}
\usepackage{float}
\usepackage{listings}
\usepackage{academicons}

\usepackage{tikz,xcolor}

\definecolor{lime}{HTML}{A6CE39}
\DeclareRobustCommand{\orcidicon}{
	\begin{tikzpicture}
	\draw[lime, fill=lime] (0,0) 
	circle [radius=0.16] 
	node[white] {{\fontfamily{qag}\selectfont \tiny ID}};
	\draw[white, fill=white] (-0.0625,0.095) 
	circle [radius=0.007];
	\end{tikzpicture}
	\hspace{-2mm}
}
\foreach \x in {A, ..., Z}{\expandafter\xdef\csname orcid\x\endcsname{\noexpand\href{https://orcid.org/\csname orcidauthor\x\endcsname}
			{\noexpand\orcidicon}}
}
\usepackage{caption}
\usepackage{color}
\usepackage{csquotes}
\usepackage[english]{babel}
\usepackage[affil-it]{authblk}
\usepackage[left=2.5cm,right=2.5cm, top=3cm, bottom=3cm]{geometry}
\usepackage{mathcomp}
\usepackage{breqn}
\usepackage{tabularx}
\usepackage{multirow}
\usepackage{makecell}
\usepackage{hyperref}

\usepackage[backend=biber,bibencoding=utf8, style=numeric-comp,sorting=none]{biblatex}
\bibliography{Bibliography}
\hypersetup{
    bookmarks=true,                  
    pdftoolbar=true,        
    pdfmenubar=true,       
    pdffitwindow=false,     
    pdfstartview={FitH},
    pdfauthor={Parangam Goswami, Anshuman Baruah},     
    colorlinks=true,       
    linkcolor=gray,         
    citecolor=blue,       
}
\makeatletter
\newbox\abstract@box
\renewenvironment{abstract}
  {\global\setbox\abstract@box=\vbox\bgroup
     \hsize=\textwidth\linewidth=\textwidth
    \small
    \begin{center}%
    {\bfseries \abstractname\vspace{-.5em}\vspace{\z@}}%
    \end{center}%
    \quotation}
  {\endquotation\egroup}
\expandafter\def\expandafter\@maketitle\expandafter{\@maketitle
  \ifvoid\abstract@box\else\unvbox\abstract@box\if@twocolumn\vskip1.5em\fi\fi}
\makeatother
\providecommand{\keywords}[1]{\textbf{\textit{Keywords---}} #1}
\begin{document}
\definecolor{dkgreen}{rgb}{0,0.6,0}
\definecolor{gray}{rgb}{0.5,0.5,0.5}
\definecolor{mauve}{rgb}{0.58,0,0.82}

\lstset{frame=tb,
  	language=Matlab,
  	aboveskip=3mm,
  	belowskip=3mm,
  	showstringspaces=false,
  	columns=flexible,
  	basicstyle={\small\ttfamily},
  	numbers=none,
  	numberstyle=\tiny\color{gray},
%  	keywordstyle=\color{blue},
	commentstyle=\color{dkgreen},
  	stringstyle=\color{mauve},
  	breaklines=true,
  	breakatwhitespace=true
  	tabsize=3
}
    \title{Spherically symmetric wormholes in General Relativity and modified gravity with a Kalb-Ramond background}
    % \author{Parangam Goswami\thanks{E-mail: \textsf{parangam.goswami@aus.ac.in}}}
    \author[]{Parangam Goswami\thanks{E-mail: \textsf{parangam.goswami@aus.ac.in}}\orcidA{}}
    \author[]{Anshuman Baruah\thanks{E-mail: \textsf{anshuman.baruah@aus.ac.in}}\orcidB{}}
    \author[]{Atri Deshamukhya\thanks{Correspondence to: \textsf{atri.deshamukhya@gmail.com}}\orcidC{}}
    % \author{Anshuman Baruah\orcid{0000-0001-6420-7666}}
    % \author{Atri Deshamukhya\thanks{E-mail: \textsf{atri.deshamukhya@gmail.com}}}
    \affil{Department of Physics, Assam University, Cachar - 788011, Assam, India}
    \date{}
    
    \begin{abstract}
    Among the several modified/extended gravity paradigms, the concept of antisymmetric connections leading to space-time torsion can be traced back to Cartan. More recently, developments in string theory have suggested the existence of a rank-2 self-interacting tensor field called the Kalb-Ramond field with similar outcomes, the field strength of which can support analytic wormhole-like solutions. However, detailed analyses of the physical properties of interest of such solutions are lacking. In this study, we comprehensively probe the properties of traversable Morris-Thorne like wormhole solutions sourced by the Kalb-Ramond field strength in both General Relativity (GR) and $f(R)$ and $f(R,T)$ modified gravity. We also analyze the coupling of the field strength in GR via a novel non-minimal interaction term in the action. Using suitable parametric constraints in all cases, we evaluate wormhole shape functions, numerically analyze the energy conditions near the throat, check the stability using the generalized Tolman-Oppenheimer-Volkov equation, and demonstrate the possibility of minimum exotic matter by estimating the volume integral quantifier. Our results show the existence of stable wormhole solutions in GR and a simple $f(R,T)$ gravity model, and unstable ones in a power-law type $f(R)$ gravity model.
    \\
    \\
    \keywords Wormhole, modified gravity, energy conditions, Kalb-Ramond field, TOV equation
    \end{abstract}
    
    \maketitle

    \section{Introduction}
    \label{sec:int}
    Possible modifications to Einstein’s General Relativity (GR) have been explored since the early days of the theory. Among the early approaches, Cartan \cite{cartan1923varietes,cartan1924varietes,cartan1925varietes,cartan1986manifolds} explored the concept of torsion in space-time by introducing anti-symmetric connections, in contrast to GR. Torsion is of importance in theories where space-time curvature arises from elementary particles characterised by a spin angular momentum \cite{raychaudhuri1979theoretical,hehl1976general}, and the corresponding effect on the gravitational field equations is an interesting area of study. Such a theory may realize the extension of the geometric principles of GR to physics at a microscopic level, where matter formation is characterized by a spin angular momentum in addition to the mass. However, Cartan's approach involves a drawback in that the $U(1)$ gauge invariance of the electromagnetic field cannot be preserved in this framework. This pathology can be avoided in a torsioned background with the introduction of the Chern-Simons three-form \cite{majumdar1999parity}, and interesting solutions have been reported, sparking an interest in theories with space-time torsion. More recently, advances in string-theory have shown that torsion naturally appears in the heterotic string spectrum due to the second rank anti-symmetric self-interacting tensor field, viz., the Kalb-Ramond (KR) field $B_{\mu \nu}$ \cite{kalb1974classical}. The antisymmetric three-tensor yielding the KR field strength, $H_{\mu \nu \lambda}$, can be identified as the \textit{Hodge}-dual of the derivative of the pseudoscalar \textit{axion} $H$. Static, spherically symmetric solutions of the EFEs with torsion have been reported previously, and it has been shown that the solutions can admit wormholes and naked singularities \cite{sengupta2001spherically,kar2003static}. Wormholes are exact solutions of the Einstein field equations (EFEs), describing space-time configurations with the topology being non-trivial in the interiors, and simple at the boundaries \cite{Visser:1995cc}, and in this study, we will focus on wormhole solutions in modified gravity with a KR background. Although wormholes in space-time have not been observationally verified, there are strong indications that such exotic space-times may actually be feasible, and even serve as black hole mimickers \cite{izmailov2019can,nandi2017ring}. Wormholes can be interpreted as space-time structures connecting two different asymptotically flat regions of space-time. The first traversable wormhole solutions were reported by Ellis \cite{ellis1973ether} and Bronnikov \cite{bronnikov1973scalar} independently, and next by Morris and Thorne in 1988 \cite{Morris:1988cz}. Morris and Thorne showed that such solutions can be formulated only at the expense of violating the null energy condition (NEC), requiring \textit{`exotic'} sources. Although such space-times may be supported by `ordinary' source terms, such as Maxwell and Dirac sources \cite{ konoplya2022traversable}, the NEC is always violated in GR. However, it is well-known that the violation of the NEC for wormhole sources can be minimized or completely removed in modified gravity theories, where additional curvature degrees of freedom can support such geometries \cite{PhysRevD.80.104012}. The literature on wormholes in modified gravity is vast. We will mainly be focusing on $f(R)$ and $f(R,T)$ modified gravity in this study, and some relevant works can be found in the following \cite{PhysRevD.94.044041,baruah2022new,baruah2023non,karakasis2022f,PhysRevD.80.104012,Baruah_2019,pavlovic2015wormholes,doi:10.1142/S0218271820500686,azizi2013wormhole,PhysRevD.96.044038}.
\\
As stated before, in the context of KR theory, Kar et. al \cite{kar2003static} and SenGupta \& Sur \cite{sengupta2001spherically} have demonstrated the existence of analytic wormhole-like solutions in the KR field theory due to Mazumdar and SenGupta \cite{ majumdar1999parity}. However, a detailed analysis on the properties of possible solutions is lacking in literature. In this study, we aim to probe the properties of possible wormhole solutions in KR theory in further detail. We consider the KR field in both the context of GR and modified gravity for the purpose. Here, we first consider the minimal coupling of the KR field strength to gravity in GR, and proceed to consider a novel scenario of the non-minimal coupling of the 3-form to curvature. We compute the wormhole shape functions, numerically analyze the energy conditions and stability of the space-times, and probe the possibility of the requirement of minimum exotic matter. Then, we proceed with the same analyses considering well-studied $f(R)$ and $f(R,T)$ models. Our results show that NEC-violating stable wormholes may exist in GR and the studied $f(R,T)$ model, and that such space-times may be unstable in viable $f(R)$ models. Further, we show that such space-times may possibly be supported by minimal amounts of NEC-violating matter near the throat.
\\
The remainder of the manuscript is organized as follows. In Sec. \ref{sec2} we describe the Morris-Thorne wormhole solution, set up the field equations in GR, non-minimal coupling, $f(R)$, and $f(R,T)$ gravity, and describe the geometric properties. In Sec. \ref{sec3} we present results concerning the energy conditions and stability of the space-times, and in Sec. \ref{sec4} we present relevant discussions and conclude the work. We adhere to the natural system of units ($G=c=1$) throughout the work.

\section{The wormhole geometries}\label{sec2}
Wormholes can be obtained as exact solutions to the EFEs in GR and modified gravity. In this study, we are interested in static, spherically symmetric space-times described by a line element of the form
\begin{equation}
    \label{mtle}
     ds^2 = - e^{2 \Phi(r)} dt^2 + \frac{dr^2}{ 1 - \frac{b(r)}{r}} + r^2 {d{\theta}}^2 + r^2 sin^2 {\theta} d {\phi}^2
    \end{equation}
This metric ansatz was first used by Morris \& Thorne \cite{Morris:1988cz}. To avoid singularities, the proper radial coordinate $l(r) = \int_{r_0}^r \frac{dr}{\sqrt{1-\frac{b}{r}}}$ should be well behaved, which imposes the restriction $\frac{b}{r} \leq 1$. It turns out that this space--time is a special case of the Ellis--Bronnikov \cite{ellis1973ether,bronnikov1973scalar} space--time described in terms of $l(r)$ as
\begin{equation}
    \label{Ellis_B_1}
     ds^2 = - dt^2 e^{2 \Phi (l)} + dl^2+ r^2(l) ({d{\theta}}^2 + r^2 sin^2 {\theta} d {\phi}^2)
    \end{equation}
where the throat is located at the minimum of $r(l)$.

    The first coefficient of the line element in Eq. \eqref{mtle} gives a measure of the gravitational redshift, and $\Phi(r)$ is known as the redshift function. The function $b(r)$ in the second coefficient of the line element in Eq. \eqref{mtle} determines the topological configuration of the space-time, and is referred to as the shape function. The throat of the wormhole is located at some value $r_{0}$ of the radial co-ordinate $r$. Moreover, traversability requires that the throat is not surrounded by an event horizon. In spherically symmetric space-times, horizons are identified as physically non-singular surfaces at $g_{00}=-e^{2\Phi} \rightarrow 0$, leading to the constraint that $\Phi(r)$ should be well defined throughout the space-time. The condition for traversability demands the following geometric constraints on the shape function: (i) $b(r_o) = r_o$, (ii) $\frac{b(r) - b^{\prime} (r) r}{b^2} > 0$, (iii) $b^{\prime} (r_o) - 1 \leq 0$, (iv) $\frac{b(r)}{r} < 1, \forall r > r_o$, (v) $\frac{b(r)}{r} \rightarrow 0$ as $r \rightarrow \infty$, where prime denotes a derivative with respect to the radial co-ordinate $r$. These constraints on the metric functions constrain the energy density $\rho$, radial $p_{r}$, and transverse $p_{t}$ pressures of the matter sources through the EFEs. Therefore, owing to these constraints, violations of the energy conditions appear while constructing traversable wormhole configurations. Here, we wish to study if $b(r)$ obtained from the field equations of the KR--gravity models satisfy these constraints (i.e., describe traversable Lorentzian wormholes) with the presence of the KR field strength. As the viability for the shape function can be well understood in terms of the radial co-ordinate $r$, we continue the analyses with the metric ansatz \eqref{mtle}, while highlighting that one can still use the metric in terms of $l(r)$ to investigate relevant properties in both regions connected by the throat (since $l \in (-\infty, \infty)$). To probe the different properties of KR field-sourced wormholes, we will consider four cases as, \textbf{Case I:} wormhole solution with the pseudoscalar axion as the matter source in the minimal coupling (GR) scenario, \textbf{Case II:} wormhole solution with pseudoscalar axion non-minimally coupled to GR, \textbf{Case III:} wormhole solution with the pseudoscalar axion as the matter source in $f(R)$ gravity, and \textbf{Case IV:} wormhole solution with the pseudoscalar axion as the matter source in $f(R,T)$ gravity. We derive the shape functions for the scenarios in this section, and present the geometric properties of the space-times.

    \subsection{Wormholes in GR (minimal coupling)}
    The field strength of the KR field is denoted by a completely antisymmetric rank three tensor $H_{\mu \nu \lambda}$. The action for the KR field strength with gravity is \cite{chakraborty2018packing}:
    
    \begin{align}
    \label{min_coup_act}
    S = \int~ d^{4} x~ \sqrt{-g} \left[~ \frac{R}{2 \kappa}~ - \frac{1}{12} H_{\mu \nu \lambda} H^{\mu \nu \lambda} \right]
    \end{align}
    
    where, $\kappa= 8 \pi$. The field equations that can be obtained from the action in Eq. \eqref{min_coup_act} are,
    
    \begin{align}
    \label{min_coup_EFE}
    G_{\mu \nu} = \kappa T_{\mu \nu} \\
    \label{axion_fe_min}
    \nabla_{\mu} H^{\mu \nu \lambda} = 0
    \end{align}
    
    where, $G_{\mu \nu} = R_{\mu \nu} - \frac{1}{2} g_{\mu \nu} R$, is the Einstein tensor, and $T_{\mu \nu}$ is the stress-energy tensor given as,
    
    \begin{align}
    \label{stress_ene_min}
    T_{\mu \nu} = \frac{1}{6} \left[ 3 H_{\mu \alpha \beta} {H_{\nu}}^{\alpha \beta} - \frac{1}{2} g_{\mu \nu} H_{\alpha \beta \gamma} H^{\alpha \beta \gamma} \right]  
    \end{align}
    
    The dual psuedoscalar ``axion" $H$ can be defined as,
    
    \begin{align}
    \label{axion_scalar}
    H_{\mu \nu \lambda} = \epsilon^{\sigma}_{\mu \nu \lambda} \partial_{\sigma} H
    \end{align}
    
    For the KR field strength, the Bianchi identity is,
    
    \begin{align}
    \label{bianchi_axion}
    \epsilon^{\mu \nu \lambda \sigma} \partial_{\sigma} H_{\mu \nu \lambda} = 0
    \end{align}
    
    In four dimensions, the KR field possess only one degree of freedom \cite{maity2004parity,sengupta2001spherically,kar2003static,chakraborty2018packing}. In the background of the spherically symmetric line element described in Eq. \eqref{mtle}, the only surviving non-zero component is $H_{023}$, and it depends only on the radial co-ordinate $r$. Therefore, the square of the field strength can be written as $H_{\mu \nu \lambda} H^{\mu \nu \lambda} = 6 H_{023} H^{023} = {[h(r)]}^2$ \cite{chakraborty2018packing,cox2016stability}. Using the field equation in Eq. \eqref{axion_fe_min}, the Bianchi identity in Eq. \eqref{bianchi_axion}, and the line element in Eq. \eqref{mtle}, we get the relation
    
    \begin{align}
    \label{num_value_h}
    \partial_{1} \left( e^{\Phi} r^{3/2} {(r-b)}^{1/2} H' \right) = 0,
    \end{align}
    
    where, a prime denotes derivative with respect to $r$. Eq. \eqref{num_value_h} can be solved to obtain
    
    \begin{align}
    \label{h_value}
    h(r) = H'(r) {(r-b)}^{1/2} = \frac{B}{r^{3/2}} e^{-\Phi}
    \end{align}
    
    where $B$ is a constant of integration. Moreover, the components of the stress-energy tensor from Eq. \eqref{stress_ene_min} can be written as 
    
    \begin{align}
    \label{st_compo_min}
    T^{0}_{0}&= - {h(r)}^2 = - \rho \nonumber \\
    T^{1}_{1}&= {h(r)}^2 = - T^{2}_{2} = - T^{3}_{3} = p
    \end{align}
    
    In the context of spherical symmetry, Eq. \eqref{st_compo_min} suggests that the stress-energy tensor of the KR field can be expressed as a perfect fluid, having the diagonal stress-energy tensor \cite{chakraborty2018packing}.
    \\
    As described previously, the parameter $\Phi(r)$ in Eq. \eqref{mtle}, provides a measure of the gravitational redshift and should be well defined through out the space-time. However, there remains the possibility of a simple form of solution known as the zero-tidal force solution, where $\Phi ’(r)=0$ everywhere \cite{Morris:1988cz}. Stationary observes shall measure precisely zero-tidal forces, and this particular form of the solution is termed in literature as the ``tideless" solution. Moreover, setting $\Phi ’(r) = 0$, simplifies the calculations significantly. 
    
    For an anisotropic $T_{\mu \nu}$, with the background line element in Eq. \eqref{mtle}, the EFEs are,
    
    \begin{align}
    \label{MT_min_coup_1}
    \frac{b'}{r^2} = \kappa \left(\rho \right) \\
    \label{MT_min_coup_2}
    - \frac{b}{r^3} = \kappa \left( p_{r} \right) \\
    \label{MT_min_coup_3}
    \left(1-\frac{b}{r}\right) \left(- \frac{b' r - b}{2 r^2 (r-b)} \right) = \kappa \left( p_{t} \right) 
    \end{align}
    
    The Ricci scalar is given as $R=\frac{2b'}{r^2}$. From the line element in Eq. \eqref{mtle}, we get $h^2= \frac{B^2}{r^3}$ and using Eq. \eqref{MT_min_coup_1}, we get,
    
    \begin{align}
    \label{int_sf}
    b = \kappa \int \frac{B^2}{r} dr
    \end{align}
    
    Solving Eq. \eqref{int_sf}, we get,
    
    \begin{align}
    \label{const_sf}
    b = \kappa B^2 \log (r) + c_{1}
    \end{align}
    
    where, $c_{1}$ is a constant of integration. Eq. \eqref{const_sf} gives the form of the shape function for the wormhole the minimally coupled scenario, and we use the same approach for the remainder of the cases as well. In order to determine $c_{1}$, we use one of the constraints on the shape function, viz. $b(r_0) = r_0$. Therefore, the shape function becomes
    
    \begin{align}
    \label{min_sf}
    b = r_0 + \kappa B^2 \log(\frac{r}{r_0})
    \end{align}
    
\begin{figure}[H]
\includegraphics[width=\textwidth]{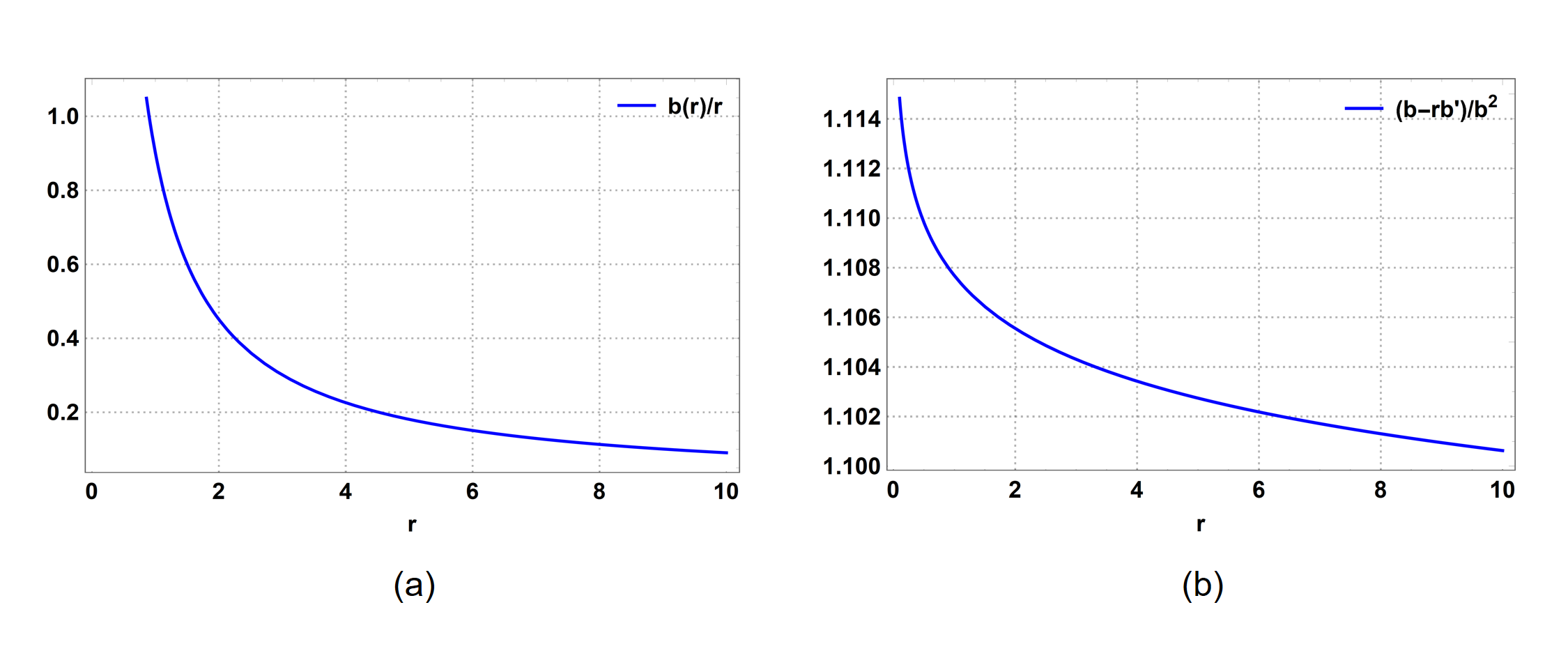}
\caption{Properties of the shape function. Profile of the (a) asymptotic flatness $\frac{b}{r}$ vs. $r$ and (b) flaring out condition $\frac{b-rb'}{b^2}$ vs. $r$ with $r_0=0.9$, and $B=0.01$ for Case I.\label{fig:one}}
\end{figure}

    The profiles of the shape function have been shown in Figure \ref{fig:one}, and it can be seen that the space-time is asymptotically flat, and flares out for $B=0.01$. With this particular form of the shape function, the various energy conditions and stability are numerically analyzed and presented in the next sections.

\subsection{Wormholes with non-minimal coupling of the KR field}
\label{sec:nonmin}

	In the context of the non-minimal coupling of the KR field to gravity, the well-studied paradigm involves the non-minimal coupling of the KR vacuum expectation value (VEV) with curvature. Although the KR Lagrangian is Lorentz-invariant, coupling the KR-VEV non-minimally yields the peculiar property of spontaneously breaking of Lorentz symmetry locally \cite{kostelecky1989spontaneous, altschul2010lorentz}, while preserving the $U(1)$ gauge invariance of the electromagnetic theory in a torsioned background. Wormhole solutions have been analyzed in this framework before \cite{lessa2021traversable}. Here, we deviate from this approach, and present a novel action for the KR field strength non-minimally coupled to gravity instead:

    \begin{align}
    \label{non_min_act_1}
    S = \int~ d^{4} x~ \sqrt{-g} \left[~ \frac{R}{2 \kappa}~ - \frac{1}{12} H_{\mu \nu \lambda} H^{\mu \nu \lambda}~ - \frac{\xi}{12} R H_{\mu \nu \lambda} H^{\mu \nu \lambda}~ \right]
    \end{align}
    
    where, $\kappa=8 \pi$, and $\xi$ is the non-minimal coupling constant. For simplicity, only the first order non-minimal coupling is considered, i.e., coupling with the Ricci scalar $R$. The field equations obtained from the action in Eq. \eqref{non_min_act_1} are,
    
    \begin{align}
    \label{non_min_efe}
    G_{\mu\nu}=\kappa {\left[ T^{eff} \right]}_{\mu\nu} \\
    \label{axion_non_min_fe}
    \nabla_{\mu} \left[ (1+ \xi R) H^{\mu \nu \lambda} \right] = 0
    \end{align}
    
    where ${\left[ T^{eff} \right]}_{\mu\nu}$ is the effective stress-energy tensor given as
    
    \begin{align}
    \label{eff_stress}
    {\left[ T^{eff} \right]}_{\mu\nu}&~ =~ \frac{1}{6-\kappa\xi (H_{\alpha\beta \gamma}H^{\alpha\beta\gamma})}~ \left[ 3 H_{\mu\alpha\beta}{H^{\alpha\beta}_{\nu}}~ -~ \frac{1}{2}g_{\mu\nu}H_{\alpha\beta \gamma}H^{\alpha\beta\gamma} \right. \nonumber\\ 
    &+~ \left. \xi~ \left\{ 3 R  H_{\mu\alpha \beta} H^{\alpha\beta}_{\nu}~ -~  \nabla_{\mu}\nabla_{\nu} \left( H_{\alpha\beta \gamma}H^{\alpha\beta\gamma} \right)~ +~ g_{\mu\nu} \Box \left( H_{\alpha\beta \gamma}H^{\alpha\beta\gamma} \right) \right\} \right]
    \end{align}
    
    With the components,
    
    \begin{align}
    \label{st_compo_non_min}
    [T^{eff}]^{0}_{0}&= - \frac{1}{1-\kappa \xi (h^2)} \left\{ h^2 \left(1+3\xi R \right) \right\} = - \rho^{eff} \nonumber \\
    [T^{eff}]^{1}_{1}&= \frac{1}{1-\kappa \xi (h^2)} \left\{ h^2 \left(1+3\xi R \right) \right\} = - [T^{eff}]^{2}_{2} = - [T^{eff}]^{3}_{3} = p^{eff}
    \end{align}
    
    Finally, the field equations are obtained as
    
    \begin{align}
    \label{fe_non_min_1}
    \frac{b'}{r^2} = \kappa \left( {\rho}^{eff} \right)\\
    \label{fe_non_min_2}
    - \frac{b}{r^3} = \kappa \left( p_r^{eff} \right)\\
    \label{fe_non_min_3}
   \left(1-\frac{b}{r}\right) \left(- \frac{b' r - b}{2 r^2 (r-b)} \right) = \kappa  \left( p_t^{eff} \right)
    \end{align}
    
    The Ricci scalar for the line element in Eq. \eqref{mtle} is given as $R=\frac{2 b'}{r^2}$, and $h^2= \frac{B^2}{r^3}$ here. Now, from Eq. \eqref{fe_non_min_1}, we get
    
    \begin{align}
    \label{non_min_sf_int}
    b=\int \frac{B^2 \kappa  r^4}{5 B^2 \kappa  \xi +r^3} \, dr
    \end{align}
    
    Solving Eq. \eqref{non_min_sf_int}, we get,
    
    \begin{align}
    \label{non_min_sf_c}
    b=&\frac{1}{6} B^2 \kappa  \left[ -5^{2/3} B^{4/3} ({\kappa \xi})^{2/3} \log \left(5 B^{4/3} ({\kappa \xi})^{2/3} - {(5 B)}^{2/3} \sqrt[3]{\kappa \xi}~ r + \sqrt[3]{5}~ r^2\right) \right. \nonumber \\
    &+ \left. 2\ 5^{2/3} B^{4/3} ({\kappa \xi})^{2/3} \log \left(5 B^{2/3} \sqrt[3]{\kappa \xi} + 5^{2/3}~ r\right) \right. \nonumber \\
    &- \left. 2 \sqrt{3} 5^{2/3} B^{4/3} {(\kappa \xi)}^{2/3} \tan ^{-1}\left(\frac{\frac{2\ 5^{2/3} r}{B^{2/3} \sqrt[3]{\kappa \xi}}-5}{5 \sqrt{3}}\right) + 3 r^2\right] + c_{2}
    \end{align}
    
    where, $c_{2}$ is a constant of integration. The constant $c_{2}$ is analysed using the condition at the wormhole throat, $b(r_0) = r_0$. Thus, the shape function then takes the form
    
    \begin{align}
    \label{final_sf_non_min}
    b=& \frac{1}{6} 5^{2/3} B^{10/3} \kappa ^{5/3} \xi ^{2/3} \left[\log \left(5 B^{4/3} {(\kappa \xi)}^{2/3} - {(5 B)}^{2/3} \sqrt[3]{\kappa \xi}~ r_0 + \sqrt[3]{5}~ r_0^2\right) \right. \nonumber \\
    &- \left. 2 \log \left(5 B^{2/3} \sqrt[3]{\kappa \xi} + 5^{2/3}~ r_0\right)+2 \sqrt{3} \tan ^{-1}\left(\frac{\frac{2\ 5^{2/3} r_0}{B^{2/3} \sqrt[3]{\kappa \xi} }-5}{5 \sqrt{3}}\right) \right. \nonumber \\
    &- \left. \log \left(5 B^{4/3} {(\kappa \xi)}^{2/3} - {(5 B)}^{2/3} \sqrt[3]{\kappa \xi }~ r + \sqrt[3]{5}~ r^2\right) + 2 \log \left(5 B^{2/3} \sqrt[3]{\kappa \xi} \right. \right. \nonumber \\
    &+ \left. \left. 5^{2/3}~ r\right) + 2 \sqrt{3} \tan ^{-1}\left(\frac{5-\frac{2\ 5^{2/3} r}{B^{2/3} \sqrt[3]{\kappa \xi}}}{5 \sqrt{3}}\right)\right] + \frac{1}{2} B^2 \kappa  \left(r^2-r_0^2\right)+r_0
    \end{align}
    
\begin{figure}[H]
\includegraphics[width=\textwidth]{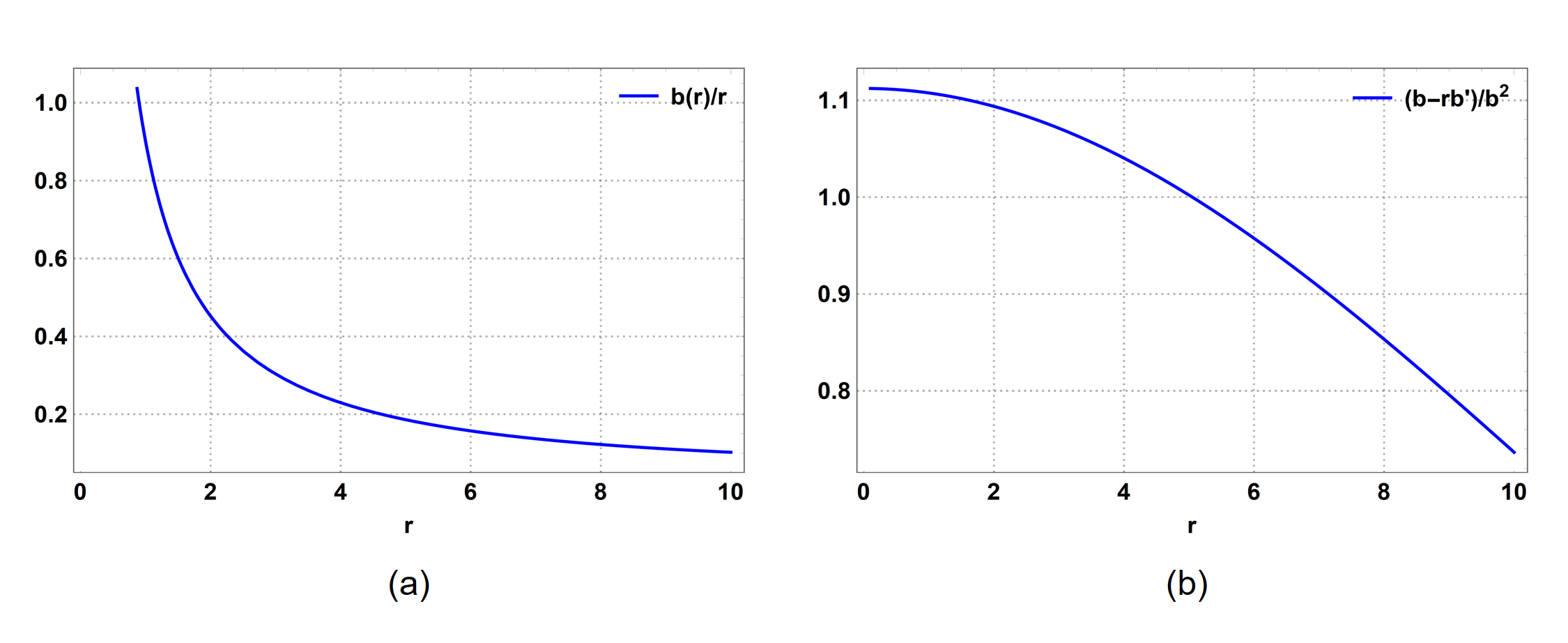}
\caption{Properties of the shape function. Profile of the (a) asymptotic flatness $\frac{b}{r}$ vs. $r$ and (b) flaring out condition $\frac{b-rb'}{b^2}$ vs. $r$ with $r_0=0.9$, $B=0.01$, and $\xi=0.01$ for Case II.\label{fig:two}}
\end{figure}

    It is known that the non-minimal coupling constant should be small. Considering constraints from astrophysical scenarios, where a possible range is $\xi \lesssim 0.01$ \cite{sankharva2022nonminimally}, we consider the value of the non-minimal coupling parameter as $\xi = 0.01$. It can be seen from Figure \ref{fig:two} that the obtained space-time is asymptotically flat, and flares out as desired. The energy conditions and stability conditions for this space-time are presented in the next sections.
    
\subsection{Wormholes in modified gravity}
     It is well-known that several modified gravity theories can account for the shortcomings of GR such as the dark energy problem. Similarly, pathologies of stable and traversable wormholes can also be avoided in several modified gravity proposals. To probe the deviations from GR solutions in our framework, we consider the case of $f(R)$ and $f(R,T)$ modified gravity theories, which serve as suitable alternatives to GR in resolving several issues. Among these, $f(R)$ gravity is perhaps the most widely studied alternative to GR. In this framework, the Ricci scalar $R$ is replaced by an arbitrary function of it in the Lagrangian. Further, $f(R,T)$ gravity serves as a simple framework to non-minimal curvature-matter coupling, and is similar in its essence to the Rastall gravity framework \cite{shabani2020connection}.

        \subsubsection{$f(R)$ gravity}
        $f(R)$ modified theories are described by a general action of the form \cite{PhysRevD.80.104012},
    
    \begin{equation}
        S = \int d^4 x \sqrt{-g} \left[ f(R) + \mathcal{L}_m \right]
        \label{fract}
    \end{equation}
    
    where the symbols imply their usual meanings. The modified EFE for $f(R)$ gravity in the metric formalism can be obtained as:
    
    \begin{equation}
    FR_{\mu\nu}-\frac{1}{2}f(R)\,g_{\mu\nu}-\nabla_\mu \nabla_\nu
    F+g_{\mu\nu}\Box F=\,T^m_{\mu\nu} \,,
    \label{eq:f_R}
    \end{equation}
    
    where $F\equiv df/dR$, and $T^m_{\mu\nu} = \frac{-2}{\sqrt{-g}} \frac{\delta \mathcal{L}_m}{\delta g^{\mu \nu}}$ denotes the matter stress-energy tensor. We consider that $g_{\mu\nu}$ describes a spherically symmetric space-time described by the line-element in Eq. (\ref{mtle}), and contract Eq. (\ref{eq:f_R}) to yield the following:
    
    \begin{equation}
    FR-2f(R)+3\Box F=T
    \label{eq:f_R_box}
    \end{equation}
     
    Here, $T$ is the trace of the matter stress energy tensor and $\Box F$ is given by:
    
    \begin{align}
        \Box F= \frac{1}{\sqrt{-g}} \partial_\mu (\sqrt{-g} g^{\mu \nu} \partial_\nu F) = \left(1-\frac{b}{r}\right)\left[F''
    -\frac{b'r-b}{2r^2(1-b/r)}\,F'+\frac{2F'}{r}\right]
    \end{align}
    
    with $F'=d f(R)/d R$ and $b'=d\,b(r)/dr$. Now, substituting Eq. (\ref{eq:f_R_box}) in Eq. (\ref{eq:f_R}) yields the following modified EFEs:
    
    \begin{equation}
    G_{\mu\nu}\equiv R_{\mu\nu}-\frac{1}{2}g_{\mu\nu} R= T^{{\rm
    eff}}_{\mu\nu} \,
        \label{eq:f_R_fe}
    \end{equation}
    
    Here, $T^{{\rm eff}}_{\mu\nu}$ is an effective stress-energy tensor, generally interpreted as a gravitational fluid responsible for NEC violations. $T^{{\rm eff}}_{\mu\nu}$ comprises the matter stress energy tensor $T^m_{\mu\nu}$ and curvature stress-energy tensor $T^{{\rm c}}_{\mu\nu}$ given by:
    
    \begin{align}
    T^{c}_{\mu\nu}=\frac{1}{F}\left[\nabla_\mu \nabla_\nu F
    -\frac{1}{4}g_{\mu\nu}\left(RF+\Box F+T\right) \right]
        \label{eq:7}
    \end{align}
    
    We consider an anisotropic distribution of matter threading the wormhole geometry: 
    
    \begin{equation}
    T_{\mu\nu}=(\rho+p_t)U_\mu \, U_\nu+p_t\,
    g_{\mu\nu}+(p_r-p_t)\chi_\mu \chi_\nu \,,
    \end{equation}
    
    where $U^\mu$ is the four-velocity, and $\chi^\mu$ is a unit space-like vector.
    \\
    In the background Eq. (\ref{mtle}), the following expressions can be obtained following \cite{PhysRevD.80.104012},
    
    \begin{align}
    \rho=\frac{Fb'}{r^2}
    \label{generic1}
    \end{align}
    \begin{align}
    p_r=-\frac{bF}{r^3}+\frac{F'}{2r^2}(b'r-b)-F''\left(1-\frac{b}{r}\right)
    \label{generic2}
    \end{align}
    \begin{align}
    p_t=-\frac{F'}{r}\left(1-\frac{b}{r}\right)+\frac{F}{2r^3}(b-b'r)
    \label{generic3}
    \end{align}
    
    It is clear from the above expressions that one can easily solve for $f(R)$; however, the obtained functional may not be physically well-motivated. Thus, we leverage a well-studied cosmologically viable $f(R)$ gravity model \cite{carloni2005cosmological,nojiri2007introduction},
    
    \begin{align}
    \label{f_R_model}
    f(R) = \chi(n) R^n
    \end{align}
    
    where $n$ is a constant and $\chi(n)$ is a function of $n$. When $n=1$, $\chi(n)$ reduces to $1$, recovering GR. In order for this power law $f(R)$ gravity model to preserve the attractive nature of gravity, the parameter $\chi$ has to take positive values \cite{carloni2005cosmological,albareti2013non}.
    Using the pseudoscalar axion as the matter source as described in Eq. \eqref{st_compo_min}, Eq. \eqref{generic1} reduces to,
    
    \begin{align}
    \label{f_R_sf_int}
    b=\int \left[\frac{B^2}{r \left(\chi n \left(\frac{2}{r^2}\right)^{n-1}\right)}\right]^{1/n} \, dr
    \end{align}
    
    Solving Eq. \eqref{f_R_sf_int}, we get,
    
    \begin{align}
    \label{f_R_sf_cnst}
    b=\frac{n r}{3 (n-1)} \left(\frac{B^2 2^{1-n} \left(\frac{1}{r^2}\right)^{-n}}{\chi  n r^3}\right)^{1/n} + c_{3}
    \end{align}
    
    The constant $c_{3}$ is analyzed using the condition at the throat, $b(r_0)=r_0$. Therefore, the shape function takes the form
    
    \begin{align}
    \label{f_R_sf_final}
    b&=\frac{1}{3 (n-1)} \left[ n r \left(\frac{B^2 2^{1-n} \left(\frac{1}{r^2}\right)^{-n}}{\chi  n r^3}\right)^{1/n} \right. \nonumber \\
    &- \left. r_0 \left\{ n \left(\left(\frac{B^2 2^{1-n} \left(\frac{1}{r_0^2}\right)^{-n}}{\chi  n r_0^3}\right)^{1/n}-3\right)+3\right\} \right]
    \end{align}
    
\begin{figure}[H]
\includegraphics[width=\textwidth]{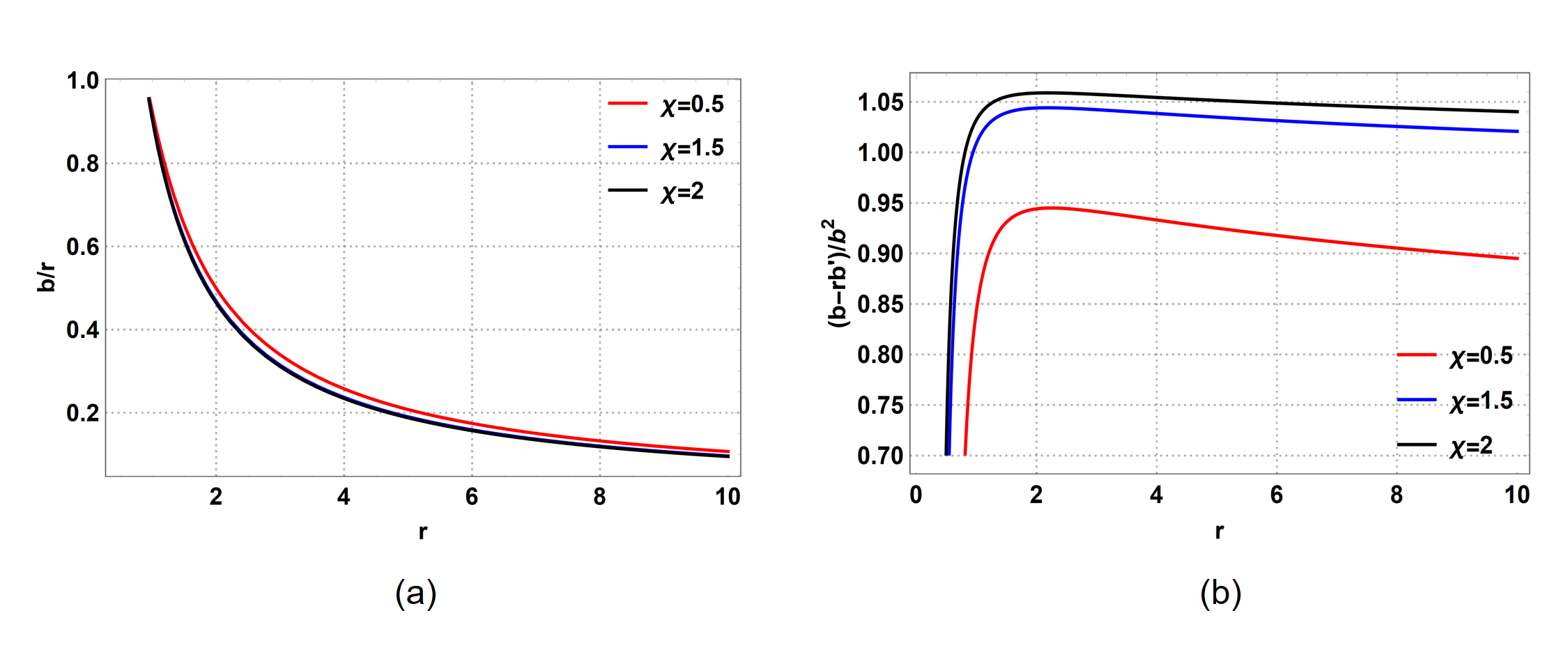}
\caption{Properties of the shape function. Profile of the (a) asymptotic flatness $\frac{b}{r}$ vs. $r$ and (b) flaring out condition $\frac{b-rb'}{b^2}$ vs. $r$ with $r_0=0.9$, $B=0.1$, and $n=1.1$ for Case III.\label{fig:three}}
\end{figure}

    The $f(R) = \chi(n) R^n$ model reduces to GR when $n=1$. Constraints on the parameters of power law gravity have been reported previously \cite{clifton2005power}. Moreover, wormhole solutions with Casimir-energy in the framework of power law gravity have been reported \cite{sokoliuk2022probing}. With these motivations and considering small deviations for GR, we have considered $n=1.1$, and $\chi=0.5, 1.5, \text{and}~ 2$. Numerical analyses of the various conditions for the viability of the shape function are shown in Figure \ref{fig:three}, and it can be seen that the obtained shape function yields a viable space-time geometry. Moreover, numerical analyses of the energy and stability conditions are reported in the next sections.

\unskip
\subsubsection{$f(R,T)$ gravity}
    $f(R,T)$ theories of gravity incorporate an extra contribution from source terms, which in turn gives rise to an extra force orthogonal to the four velocity  \cite{harko2011f}. The corrections from the trace of the stress-energy tensor, $T$, lead to several interesting outcomes such as particle production, owing to the fact that the stress-energy tensor in $f(R,T)$ gravity theory is not conserved, and the presence of anisotropic fluid in the universe \cite{harko2014thermodynamic}.
\\
    
    The action for $f(R,T)$ modified theories of gravity is given as \cite{harko2011f,PhysRevD.96.044038}
    
    \begin{equation}
    \label{f_R_T_act}
        S = \int d^4 x \sqrt{-g} \left[\frac{1}{16 \pi} f(R,T) + \mathcal{L}_m \right]
    \end{equation}
    
    Here, the Ricci scalar $R$ is replaced by the function $f(R,T)$, where $T$ is the trace of the stress-energy tensor. $\mathcal{L}_m$ is the matter Lagrangian, and is related to the stress-energy tensor as,
    
    \begin{equation}
    \label{f_R_T_st}
    T_{\mu \nu}= - \frac{2}{\sqrt{-g}}\left[\frac{\partial(\sqrt{-g}\mathcal{L}_{m})}{\partial g^{\mu \nu}}-\frac{\partial}{\partial x^{\lambda}}\frac{\partial(\sqrt{-g}\mathcal{L}_m)}{\partial(\partial g^{\mu \nu}/\partial x^{\lambda})}\right]
    \end{equation}
    
    Assuming that the $\mathcal{L}_m$ depends only on the metric tensor and not on its derivatives, and varying the action Eq. \eqref{f_R_T_st} with respect to the metric, the modified EFEs are obtained as
    
    \begin{align}
    \label{f_R_T_efe}
    &f_R(R,T) \left(R_{\mu \nu} -\frac{1}{3} R g_{\mu \nu}\right) + \frac{1}{6}f(R,T)g_{\mu \nu} = 8\pi \left(T_{\mu \nu}-\frac{1}{3}T g_{\mu \nu}\right) \nonumber \\
    &-f_T(R,T)\left(T_{\mu \nu} -\frac{1}{3}T g_{\mu \nu}\right) - f_T(R,T)\left(\theta_{\mu \nu}-\frac{1}{3}\theta g_{\mu \nu}\right) \nonumber \\
    &+\nabla_\mu \nabla_\nu f_R(R,T)
    \end{align}
    
    where, $f_R(R,T)=\partial f(R,T)/\partial R$, $f_T(R,T)=\partial f(R,T)/ \partial T$, and
    
    \begin{equation}
    \label{theta_f_R_T}
    \theta_{\mu \nu} = g^{\mu \nu}\frac{\partial T_{\mu \nu}}{\partial g^{\mu \nu}}.
    \end{equation}
    
    Considering a well-studied $f(R,T)$ model, $f(R,T)=R+2f(T)$ \cite{harko2011f}, the EFEs can be recast as 
    
    \begin{equation}
    \label{f_R_T_fe_mod}
    R_{\mu \nu}-\frac{1}{2}R g_{\mu \nu} = 8\pi T_{\mu \nu} + 2F(T) T_{\mu \nu} + [2\rho F(T)+f(T)]g_{\mu \nu},
    \end{equation}
    
    where $R_{\mu \nu}$ is the Ricci tensor and $F(T)=df(T)/dT$. Considering $f(T)=\lambda T$, where $\lambda$ is a constant, the EFEs become
    
    \begin{equation}
    \label{f_R_T_final_efe}
    G_{\mu \nu} = (8\pi+2\lambda) T_{\mu \nu} + \lambda(2\rho+T) g_{\mu \nu}
    \end{equation}
    
    Assuming that an anisotropic distribution of matter as $T^\mu_\nu = \text{diag}(-\rho, p_r, p_t,p_t)$ is threading the wormhole geometry, the components of the EFEs are
    
    \begin{align}
    \label{f_R_T_1}
    \frac{b'}{r^2}=(8\pi+\lambda)\rho-\lambda(p_r+2p_l) \\
    \label{f_R_T_2}
    -\frac{b}{r^3}= \lambda \rho+(8\pi+3\lambda)p_r+2\lambda p_l \\
    \label{f_R_T_3}
    \frac{b-b'r}{2r^3}= \lambda \rho+\lambda p_r+(8\pi+4\lambda) p_l
    \end{align}
    
    Solving Eqs. \eqref{f_R_T_1}-\eqref{f_R_T_3}, we get,
    
    \begin{align}
    \label{f_R_T_efe_1}
    \rho=\frac{b'}{r^2(8\pi+2\lambda)} \\
    \label{f_R_T_efe_2}
    p_r= - \frac{b}{r^3(8\pi+2\lambda)} \\
    \label{f_R_T_efe_3}
    p_t= \frac{b-b'r}{2r^3(8\pi+2\lambda)}
    \end{align}
    
    With the KR field strength as the matter source as described in Eq. \eqref{st_compo_min}, Eq. \eqref{f_R_T_efe_1} reduces to
    
    \begin{align}
    \label{sf_fRT_int}
    b=\int \frac{B^2 (2 \lambda +8 \pi )}{r} \, dr
    \end{align}
    
    Solving Eq. \eqref{sf_fRT_int}, we get,
    
    \begin{align}
    \label{sf_fRT_cnst}
    b = B^2 (2 \lambda +8 \pi ) \log (r) + c_{4}
    \end{align}
    
    where, $c_{4}$ is a constant of integration. The constant $c_{4}$ is analyzed using the condition at the throat, $b(r_0)=r_0$, and the shape function takes the form
    
    \begin{align}
    \label{sf_fRT_final}
    b=r_0 + 2 B^2 (\lambda +4 \pi ) \left[ \log (\frac{r}{r_0}) \right]
    \end{align}
    
\begin{figure}[H]
\includegraphics[width=\textwidth]{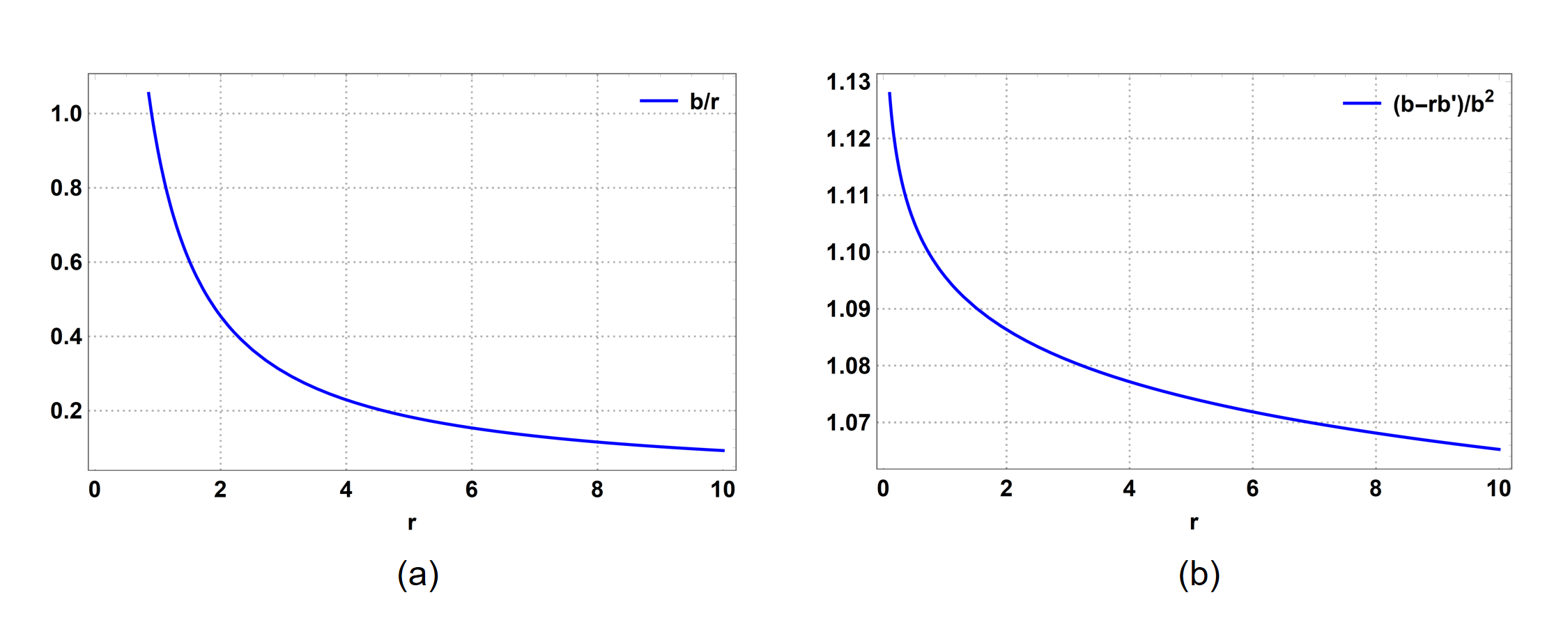}
\caption{Profile of the (a) asymptotic flatness $\frac{b}{r}$ vs. $r$ and (b) flaring out condition $\frac{b-rb'}{b^2}$ vs. $r$ with $r_0=0.9$, $B=0.1$, and $\lambda=-12$ for Case IV.\label{fig:four}}
\end{figure}

    Wormhole solutions with NEC-satisfying matter sources in the $f(R,T) = R + 2 f(T)$ model have been reported, and the parameter $\lambda$ can be constrained as $\lambda > -4 \pi$ \cite{PhysRevD.96.044038}. Considering $\lambda= -12$, the numerical analyses of the various conditions for the viability of shape function is shown in Figure \ref{fig:four}, and it can be seen that the configuration is a viable wormhole space-time. Further, the energy and stability conditions are discussed in the next sections.

\unskip

\section{Energy conditions, stability, and amount of exotic matter}
\label{sec3}
\subsection{Energy conditions}
\begin{figure}[H]
\includegraphics[width=\textwidth]{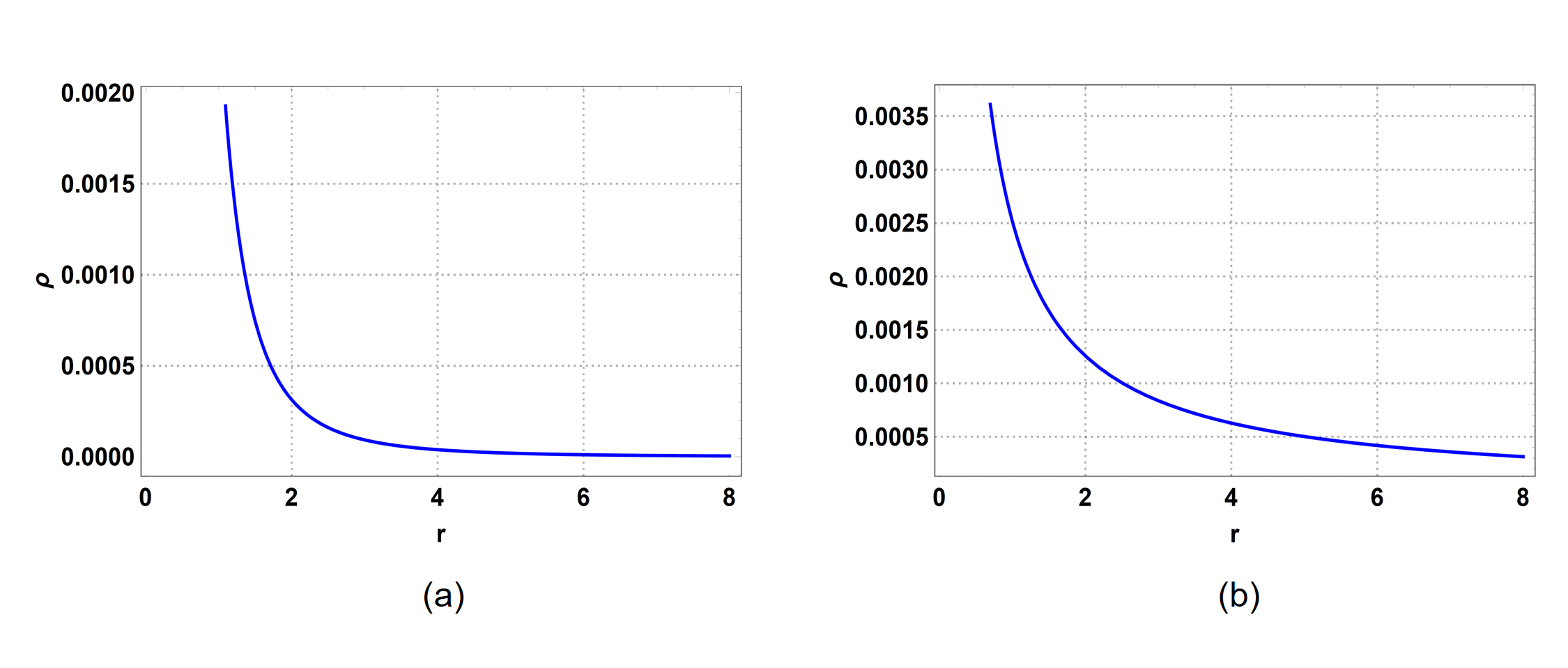}
\caption{Profile of the WEC with $r_0=0.9$, and $B=0.01$ for (a) Case I and with $\xi=0.01$ for (b) Case II.\label{fig:five}}
\end{figure}

\begin{figure}[H]
\includegraphics[width=\textwidth]{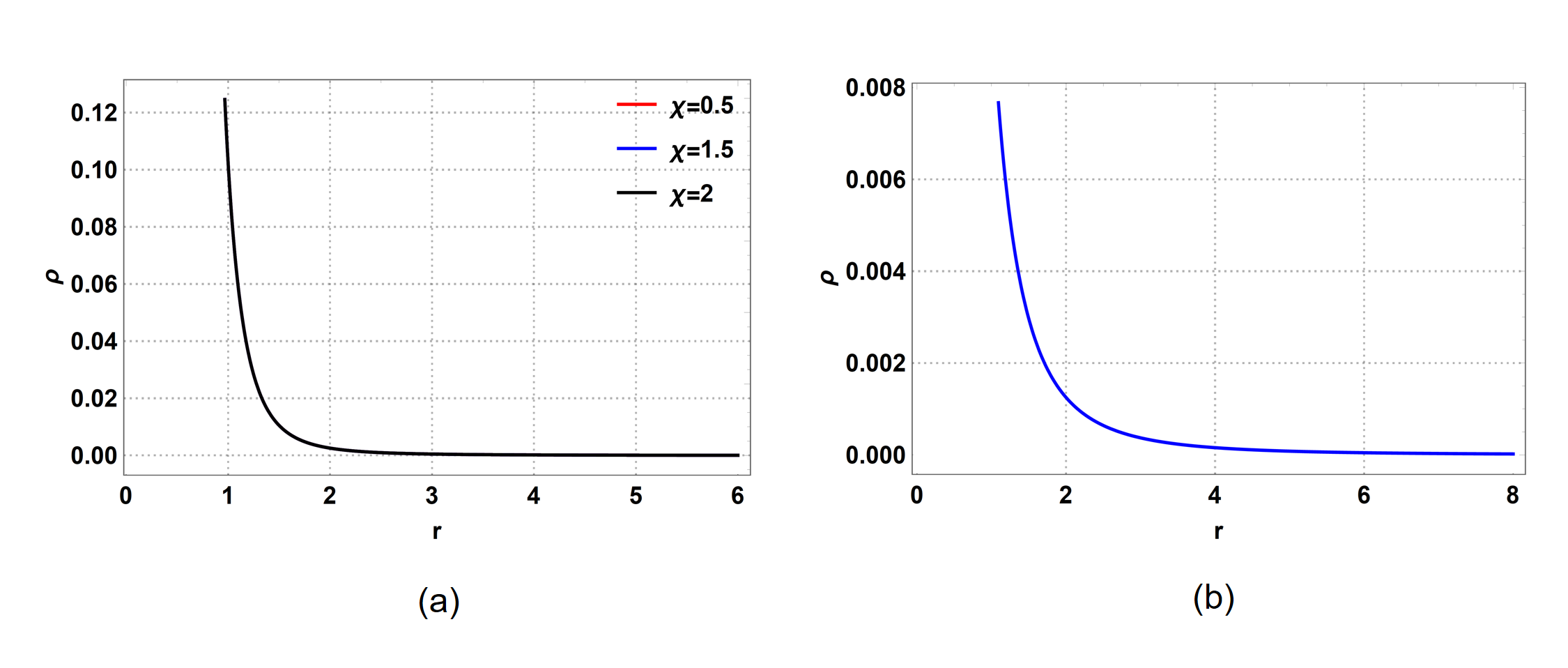}
\caption{Profile of the WEC with $r_0=0.9$, and $B=0.1$ for (a) Case III with $n=1.1$ and (b) Case IV with $\lambda=-12$.\label{fig:six}}
\end{figure}

\begin{figure}[H]
\includegraphics[width=\textwidth]{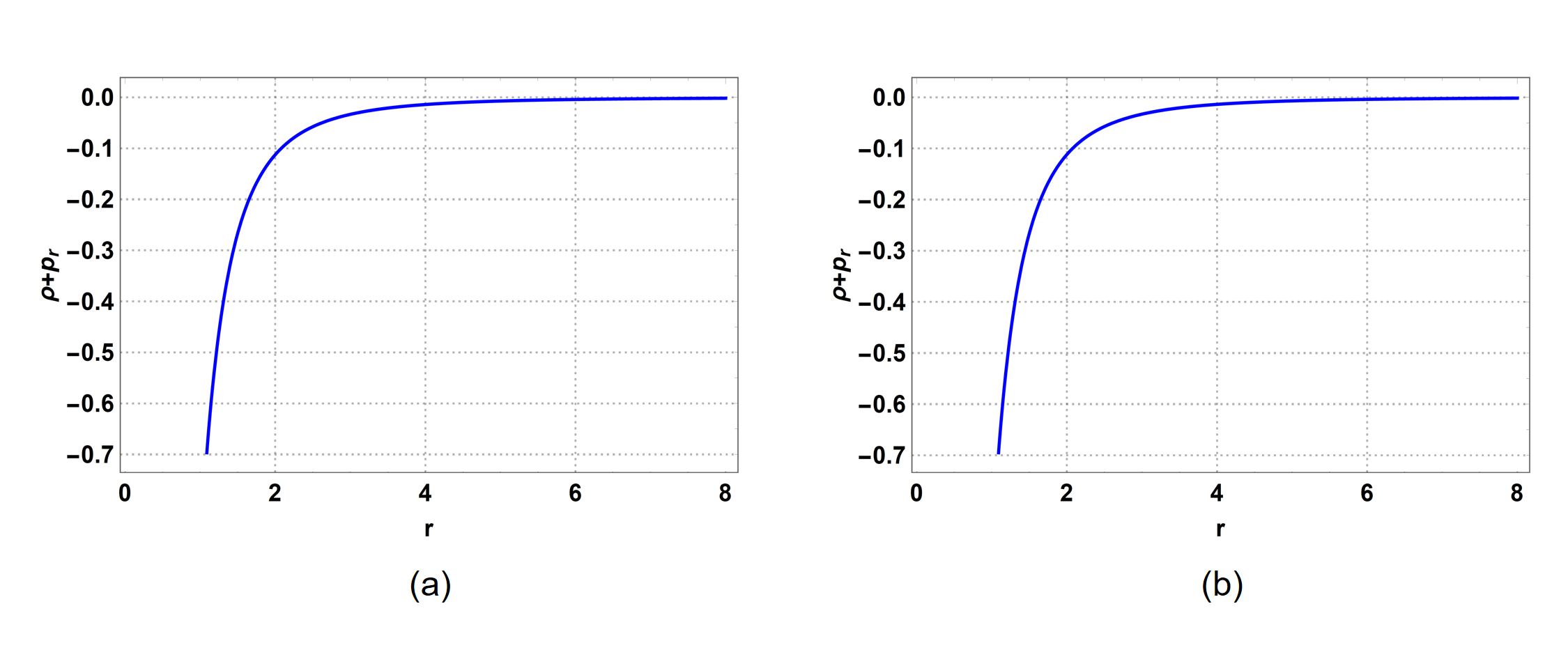}
\caption{Profile of the NEC term $\rho + p_r$ with $r_0=0.9$, and $B=0.01$ for (a) Case I and with $\xi=0.01$ for (b) Case II.\label{fig:seven}}
\end{figure}

\begin{figure}[H]
\includegraphics[width=\textwidth]{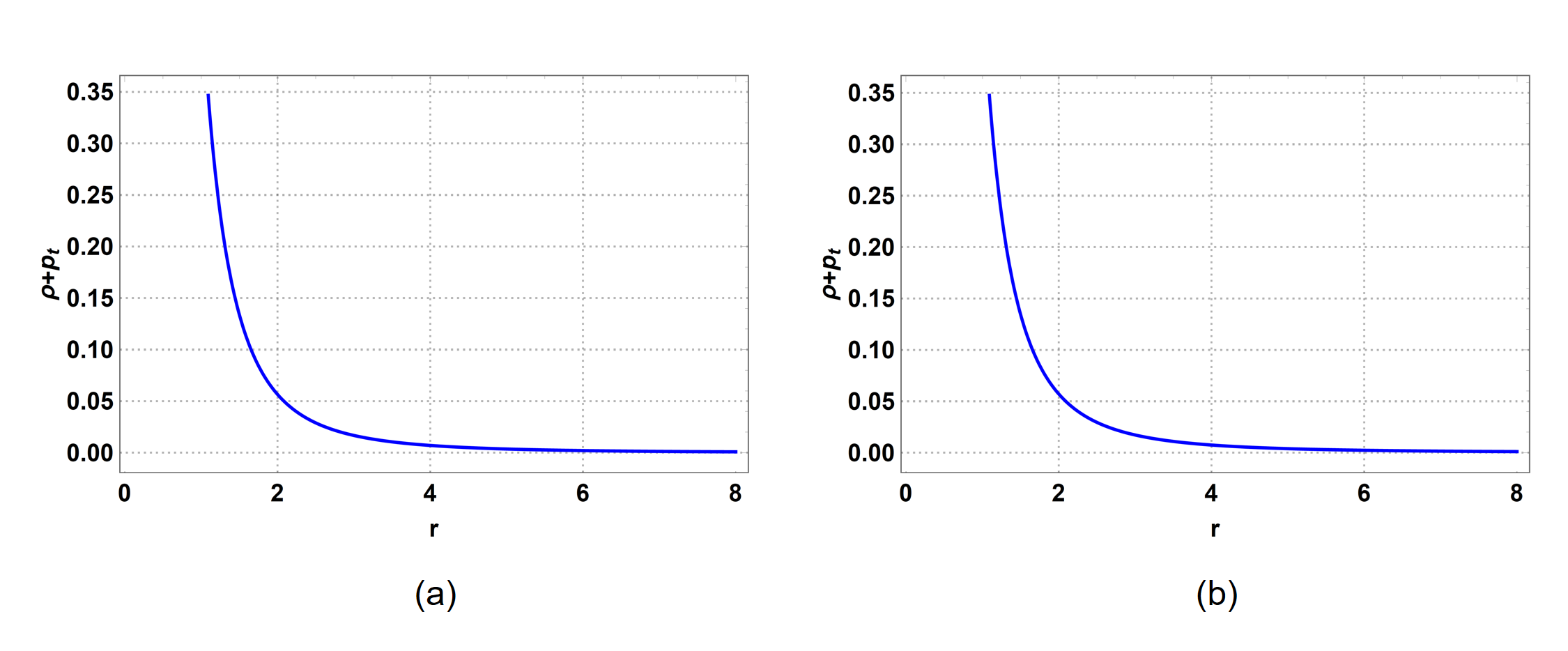}
\caption{Profile of the NEC term $\rho + p_t$ with $r_0=0.9$, and $B=0.01$ for (a) Case I and with $\xi=0.01$ for (b) Case II.\label{fig:eight}}
\end{figure}

\begin{figure}[H]
\includegraphics[width=\textwidth]{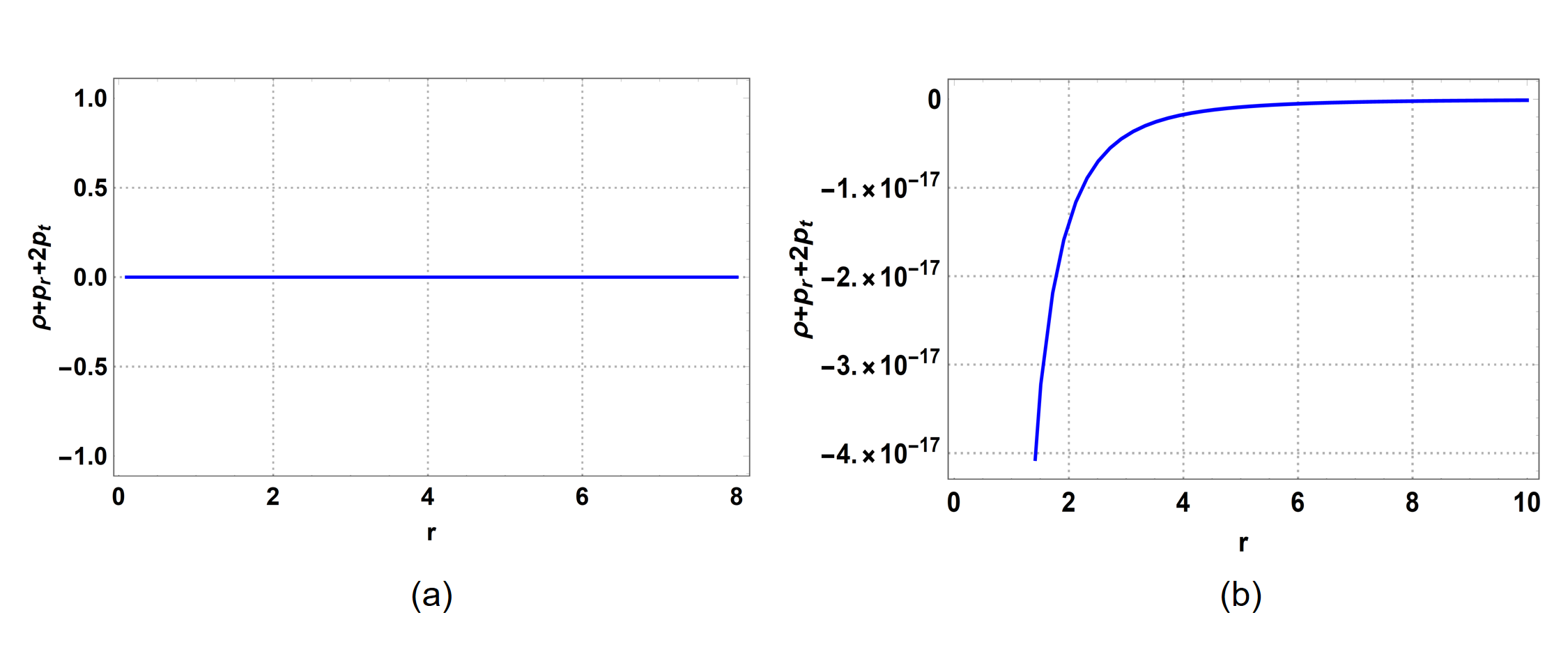}
\caption{Profile of the SEC $\rho + p_r + 2 p_t$ with $r_0=0.9$, and $B=0.01$ for (a) Case I and with $\xi=0.01$ for (b) Case II.\label{fig:eleven}}
\end{figure}

The energy conditions are sets of inequalities which the stress-energy tensor of matter sources should respect, so that the energy density of matter fields is measured to be positive by any observer traversing a time-like curve \cite{curiel2017primer}. The weak energy condition (WEC) implies $\rho \geq 0$, the NEC implies $\rho + p_r \geq 0$, and $\rho + p_t \geq 0$, the strong energy condition (SEC) implies $\rho + p_r + 2 p_t \geq 0$. These energy conditions are analyzed for all the four cases in this section.
\\
Figure \ref{fig:five} shows the WEC for Cases I and II. As expected and reported in literature \cite{chakraborty2018packing}, the pseudoscalar axion has a positive energy density and satisfies the WEC for both the cases. Figure \ref{fig:six} shows the WEC for Cases III and IV. It is seen that the KR field strength as a source shows a positive energy density, thus satisfying the WEC. In Case III, it can be seen that the energy density exhibits no apparent changes in behavior depending on the $f(R)$ model parameter $\chi$.
\\
Figure \ref{fig:seven} shows the NEC term $\rho + p_r$ for Cases I and II. It can be observed that the NEC is violated at the throat, which is a characteristic of traversable wormholes. Next, the NEC term $\rho + p_t$ is analyzed and shown in Figure \ref{fig:eight}. It can be observed that the second NEC term $\rho + p_t$ is satisfied at the throat. However, since both $\rho + p_r \geq 0$ and $\rho + p_t \geq 0$ should hold, we infer that the NEC is violated here. Further, the SEC is shown in Figure \ref{fig:eleven}.
\unskip
 
\unskip
\unskip

\begin{figure}[H]
\includegraphics[width=\textwidth]{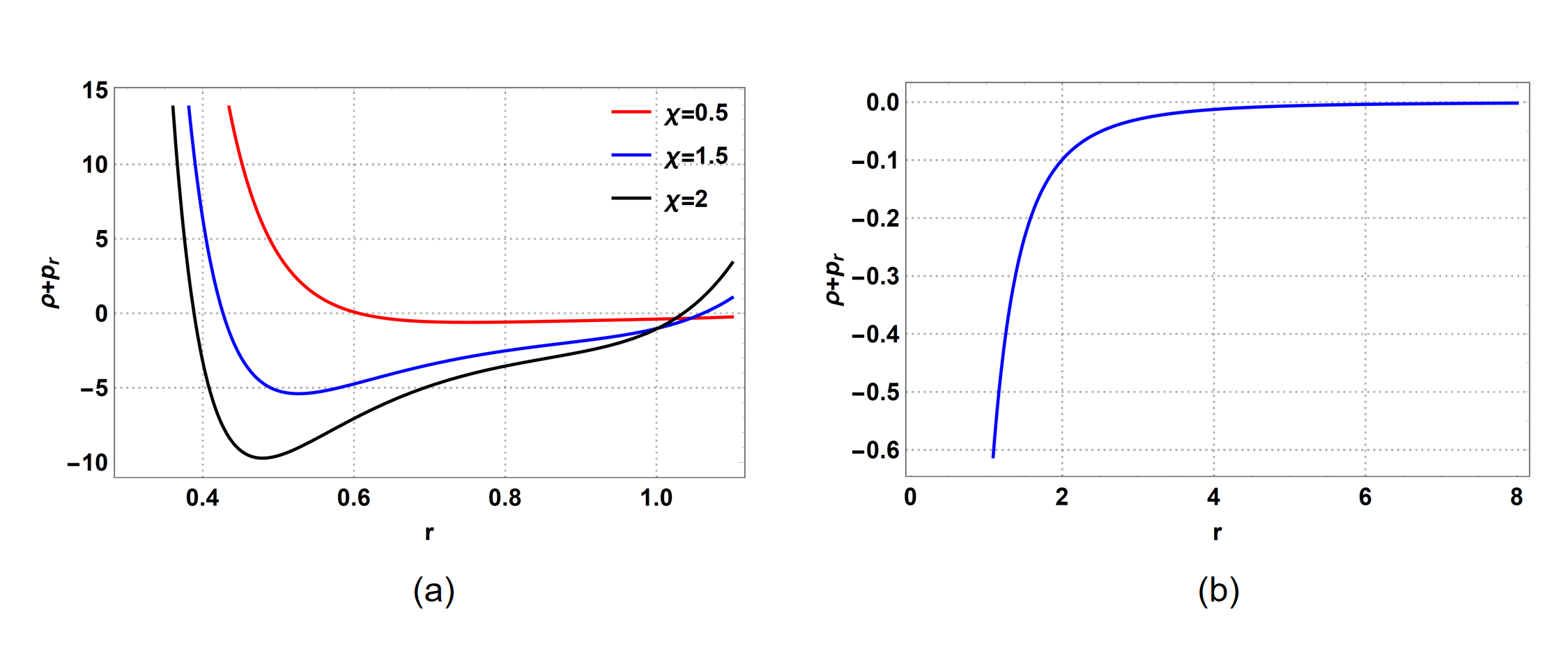}
\caption{Profile of the NEC term $\rho + p_r$ with $r_0=0.9$, and $B=0.1$ for (a) Case III with $n=1.1$ and (b) Case IV with $\lambda=-12$.\label{fig:nine}}
\end{figure}

\begin{figure}[H]
\includegraphics[width=\textwidth]{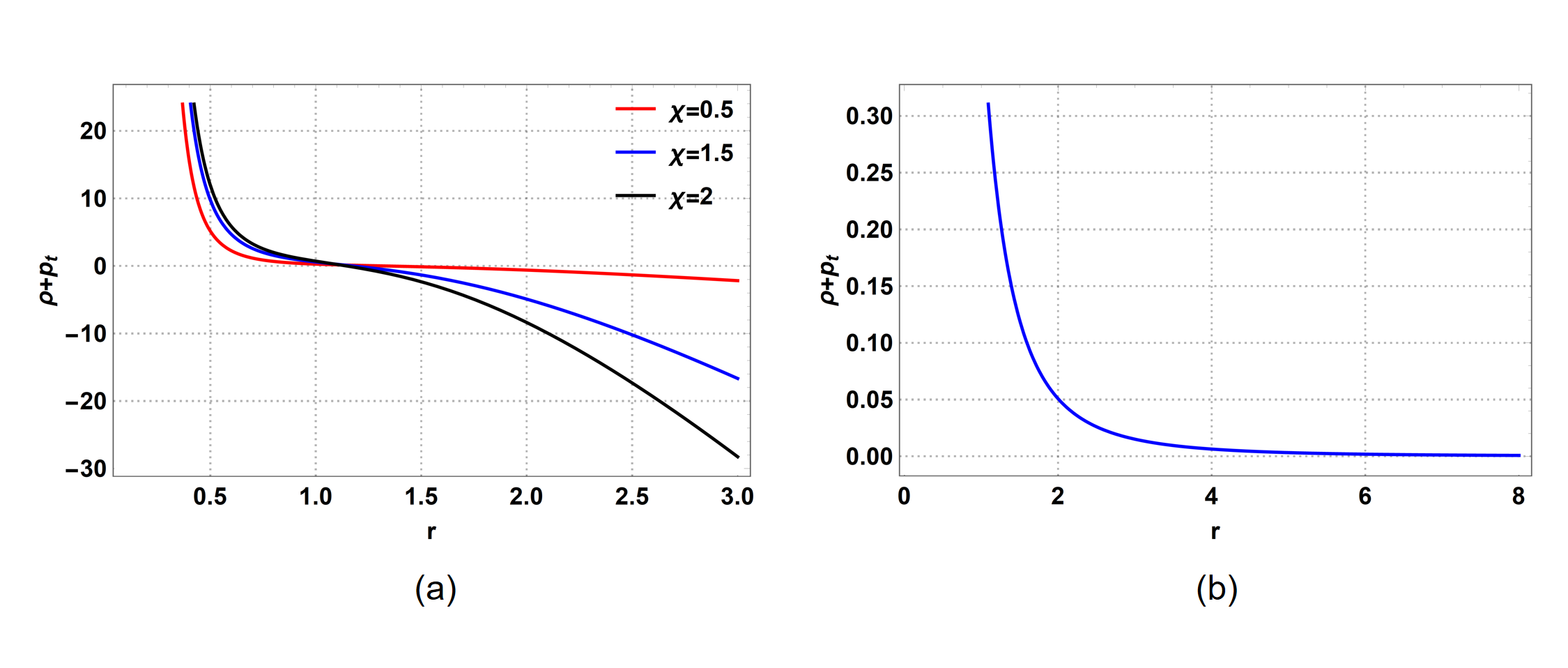}
\caption{Profile of the NEC term $\rho + p_t$ with $r_0=0.9$, and $B=0.1$ for (a) Case III with $n=1.1$ and (b) Case IV with $\lambda=-12$.\label{fig:ten}}
\end{figure}

\begin{figure}[H]
\includegraphics[width=\textwidth]{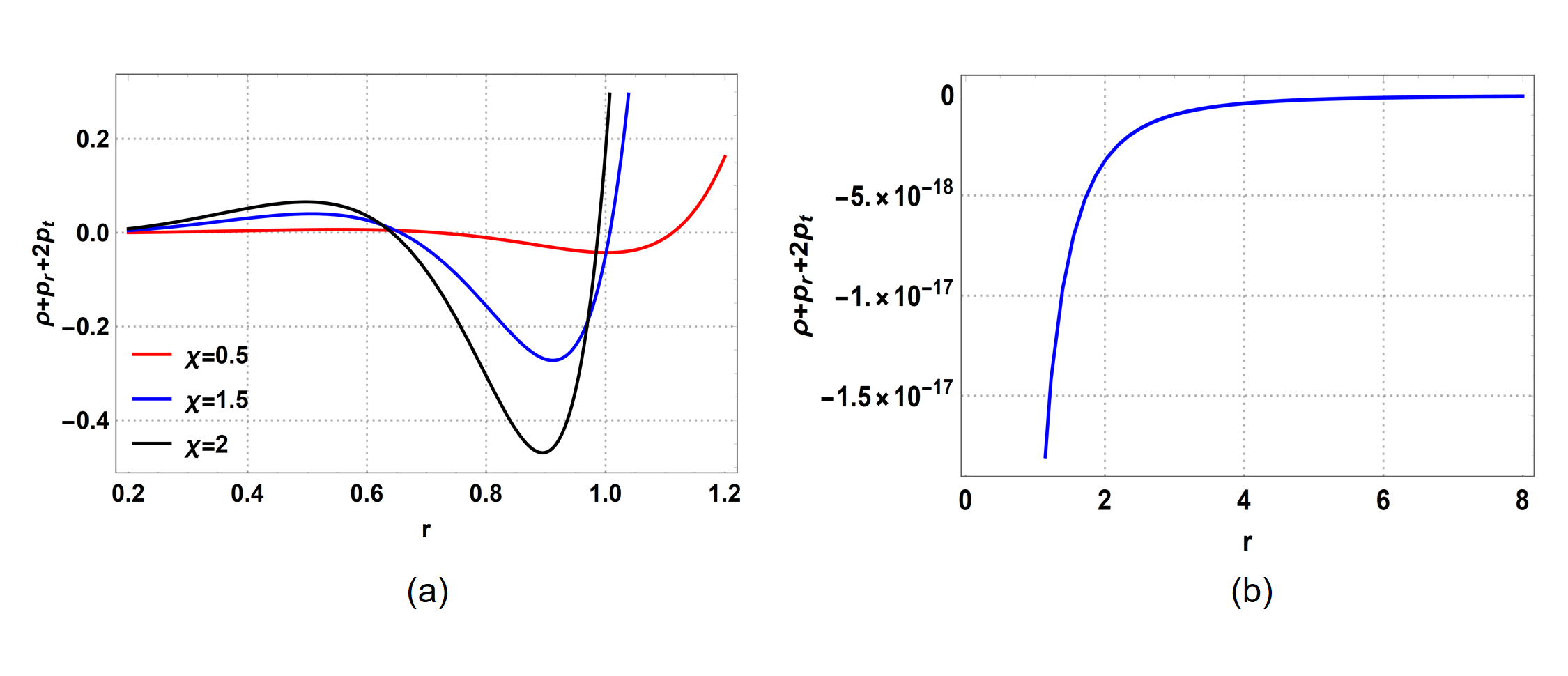}
\caption{Profile of the SEC term $\rho + p_r + 2 p_t$ vs. $r$ with $r_0=0.9$, and $B=0.1$ for (a) Case III with $n=1.1$ and (b) Case IV with $\lambda=-12$.\label{fig:twelve}}
\end{figure}

\begin{figure}[H]
\includegraphics[width=\textwidth]{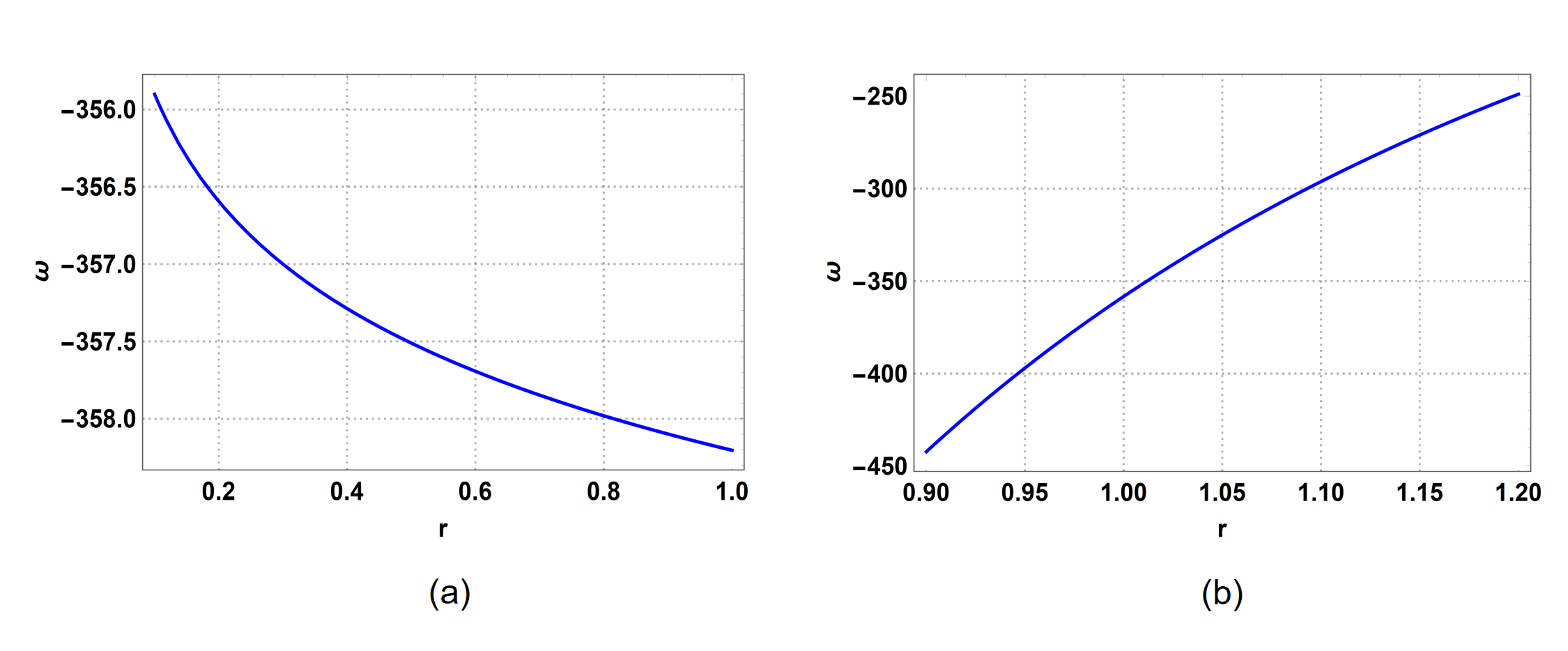}
\caption{Profile of the EoS parameter $\omega$ with $r_0=0.9$, and $B=0.01$ for (a) Case I and with $\xi=0.01$ for (b) Case II.\label{fig:thirteen}}
\end{figure}

\begin{figure}[H]
\includegraphics[width=\textwidth]{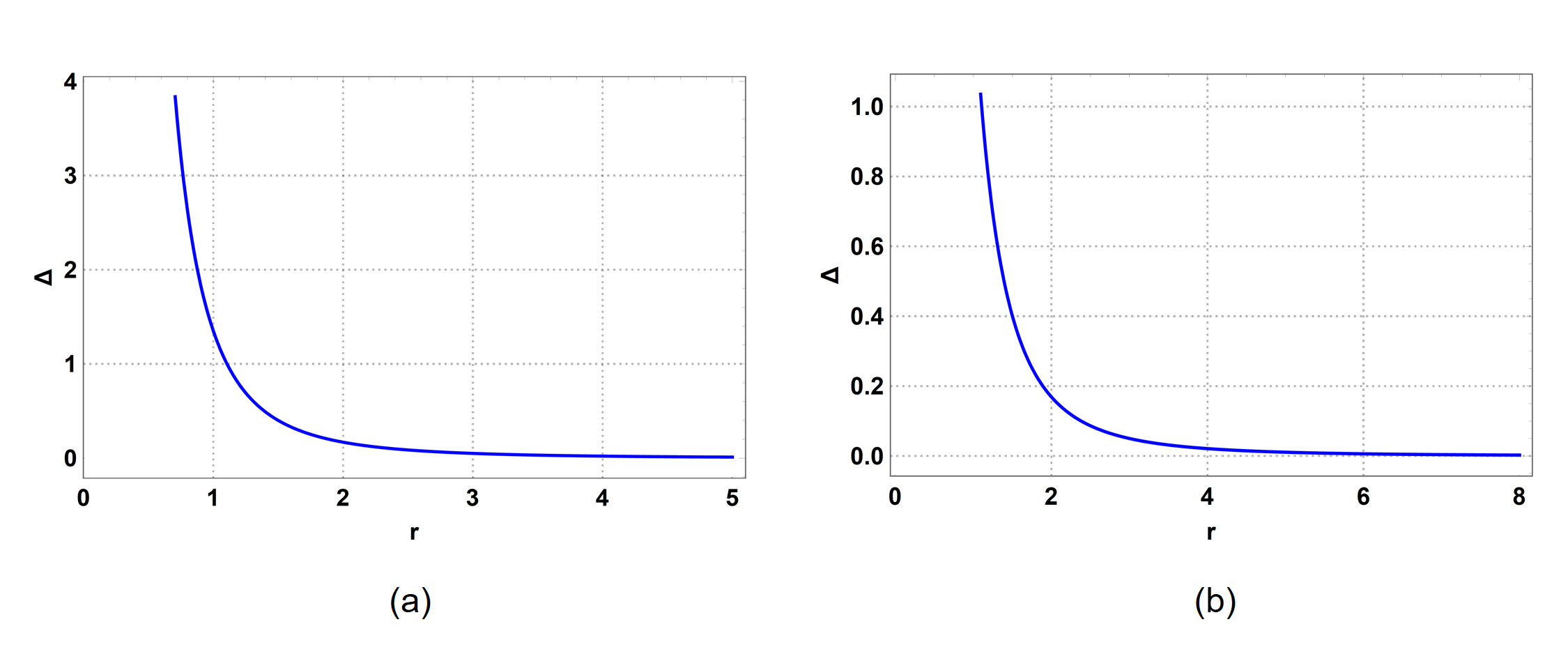}
\caption{Profile of the anisotropy parameter $\omega$ with $r_0=0.9$, and $B=0.01$ for (a) Case I and with $\xi=0.01$ for (b) Case II.\label{fig:fourteen}}
\end{figure}

\begin{figure}[H]
\includegraphics[width=\textwidth]{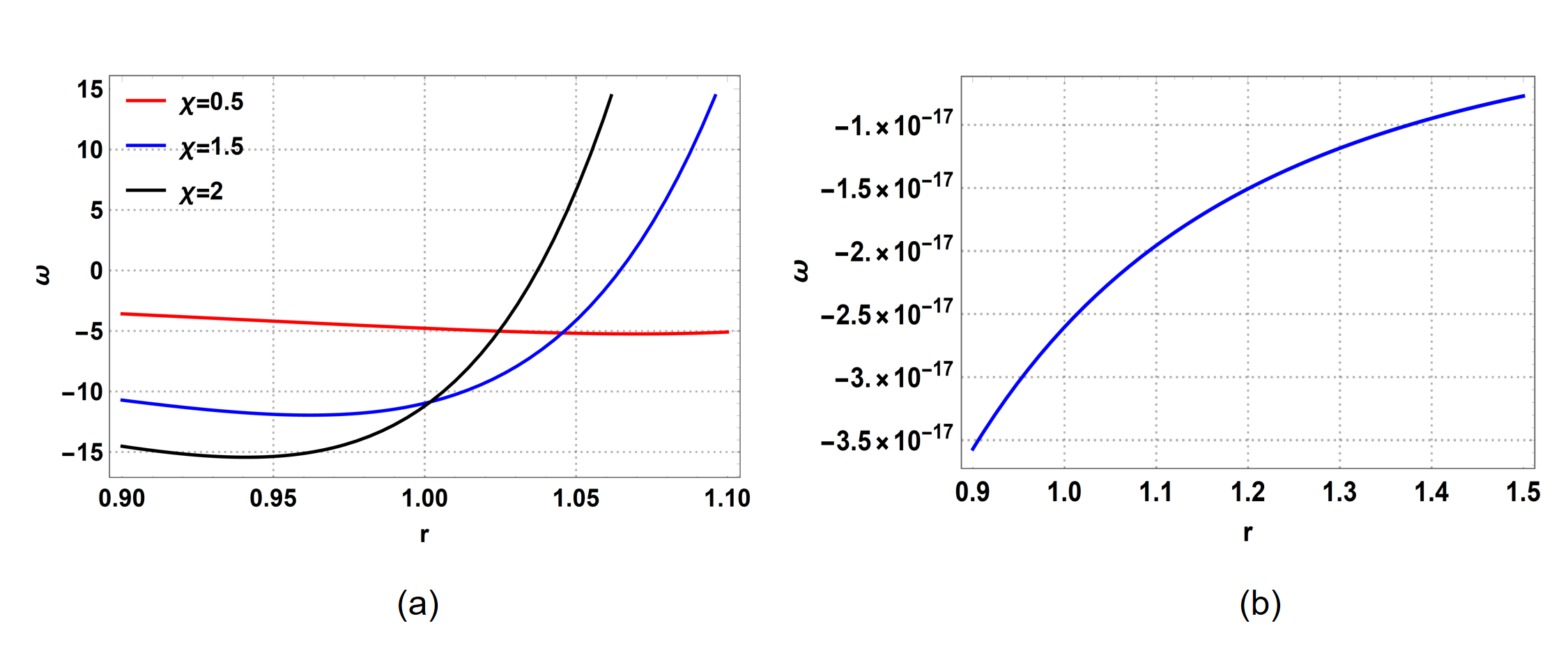}
\caption{Profile of the EoS parameter $\omega$ vs. $r$ with $r_0=0.9$, and $B=0.1$ for (a) Case III with $n=1.1$ and (b) Case IV with $\lambda=-12$.\label{fig:fifteen}}
\end{figure}

\begin{figure}[H]
\includegraphics[width=\textwidth]{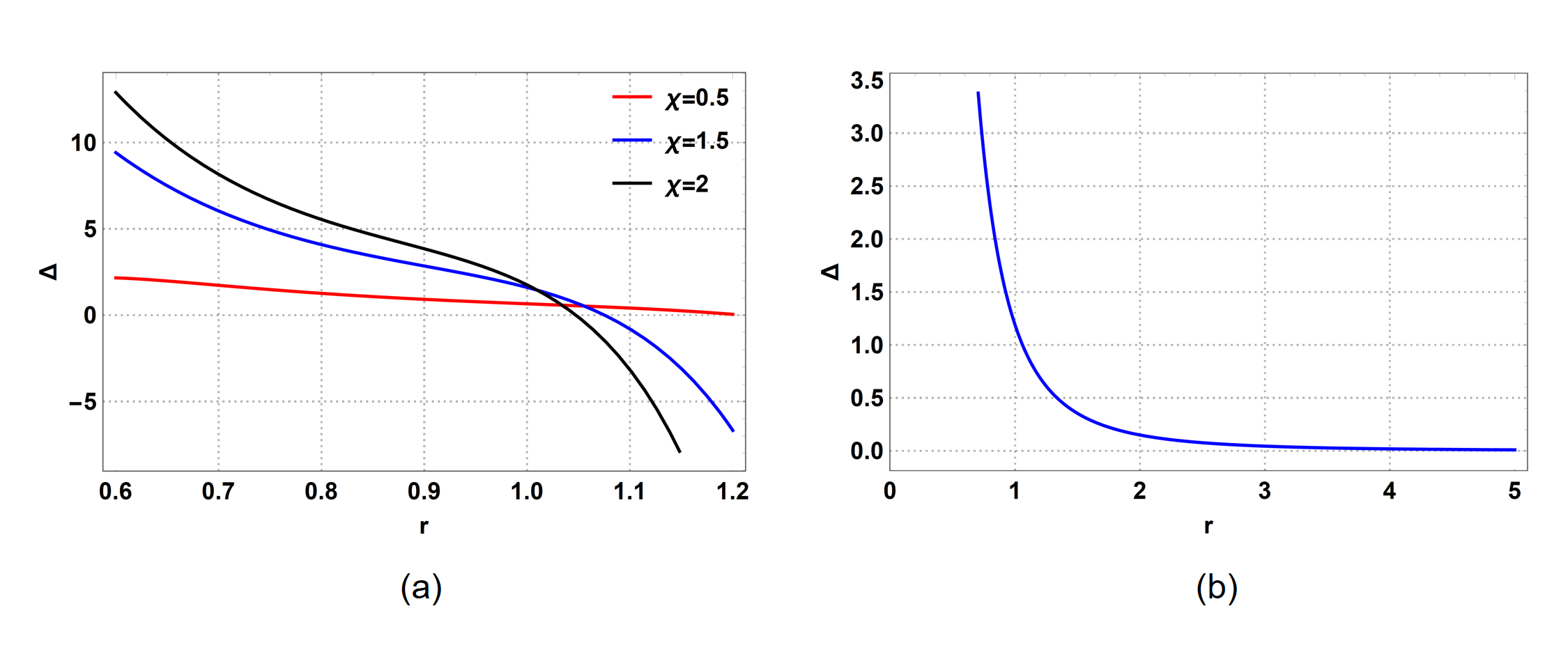}
\caption{Profile of the anisotropy parameter $\Delta$ vs. $r$ with $r_0=0.9$, and $B=0.1$ for (a) Case III with $n=1.1$ and (b) Case IV with $\lambda=-12$.\label{fig:sixteen}}
\end{figure}

\unskip
Figure \ref{fig:nine} shows the first NEC term $\rho + p_r$ for Cases III and IV. It can be observed that the first NEC term is violated at the throat for all considered values of $\chi=0.5, 1.5, \text{and}~ 2$ in Case III, and the first NEC term is also violated at the throat for Case IV with $\lambda=-12$.
\unskip
Figure \ref{fig:ten} shows the second NEC term $\rho + p_t$ for Cases III and IV. It is seen that the second NEC term is satisfied at the wormhole throat for both the cases. However, as a whole the NEC is violated for Cases III and IV owing to the violation of the first NEC term $\rho + p_r$.
\unskip
From Figure \ref{fig:eleven} it can be observed that the SEC in Case I is marginally satisfied ($\rho + p_r + 2 p_t = 0$) as reported previously \cite{chakraborty2018packing}, while the SEC for the non-minimally coupled scenario is violated at the wormhole throat.
\unskip
Figure \ref{fig:twelve} shows the SEC term $\rho + p_r + 2 p_t$ for Cases III and IV. It can be observed that the SEC is violated at the wormhole throat for both the cases.
\unskip
Along with the energy conditions, it is useful to check two more parameters namely, the equation of state (EoS) parameter $\omega = p_r/\rho$, and the anisotropy parameter $\Delta = p_t - p_r$. The EoS parameter provides information regarding the nature of the matter source threading the wormhole geometry, and the anisotropy parameter explains whether the space-time geometry is attractive or repulsive \cite{Morris:1988cz}. The profile of the EoS parameter near the throat is shown in Figure \ref{fig:thirteen}.
\unskip
From Figure \ref{fig:thirteen}, it can be observed that the EoS parameter $\omega$ has a negative value and $\omega << -1$, indicating a phantom-like behavior of the source. Moreover, it is important to note that the value of the constant ``$B$" plays a crucial role in determining these parameters. Higher values of the constant ``$B$" can lead to values of the EoS parameter $-1<\omega<0$, near the wormhole throat. However, with higher values of the constant ``$B$" the various conditions required for the viability of the shape function are not satisfied. This highlights how the metric function constraints the various energy conditions, EoS, and anisotropy parameter.
\\
\\
Figure \ref{fig:fifteen} shows the EoS parameter $\omega$ for Cases III and IV. It can be observed that the EoS parameter has a negative value near the wormhole throat for both the cases. For the case of $f(R)$ modified gravity (Case III), the EoS parameter $\omega < -1$ for $\chi = 0.5, 1.5~ \text{and}~ 2$ near the throat signifying a phantom-like nature of the source, while for the case of $f(R,T)$ modified gravity (Case IV), the EoS parameter is $-1<\omega<0$, near the wormhole throat, indicating a quintessence like behaviour of the source.
\\
Figure \ref{fig:fourteen} shows the anisotropy parameter $\Delta$ for Cases I and Case II. The anisotropy parameter is positive at the wormhole throat, signifying a repulsive geometry, as in the in-falling signals from one asymptotically flat region can cross the wormhole throat and emerge on the other region.
\\
Figure \ref{fig:sixteen} shows the anisotropy parameter $\Delta$ for Cases III and IV. The anisotropy parameter is positive near the wormhole throat for both cases, signifying a repulsive geometry, which is a characteristic feature of traversable wormholes. Table \ref{tab:one} below presents a summary of the energy conditions for Cases I-IV.
\unskip

\unskip

\unskip
\subsection{Stability and amount of exotic matter}
\begin{figure}[H]
\includegraphics[width=\textwidth]{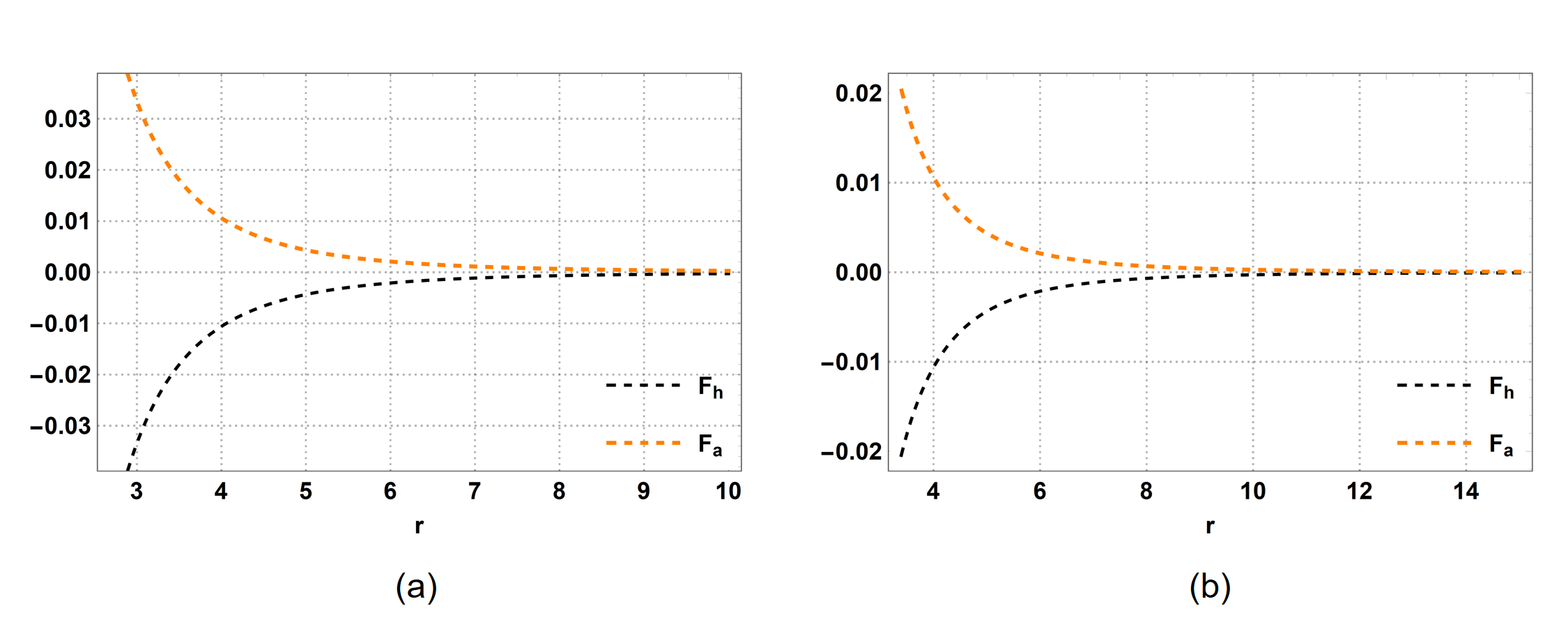}
\caption{Profile of $F_{\mathrm{h}}$ and $F_{\mathrm{a}}$ vs. $r$ with $r_0=0.9$, and $B=0.01$ for (a) Case I and with $\xi=0.01$ for (b) Case II.\label{fig:seventeen}}
\end{figure}

The stability of stellar structures can be investigated using the Tolman-Oppenheimer-Volkov (TOV) equation, first reported in the context of neutron stars \cite{tolman1987relativity,oppenheimer1939massive}. A more generalized version of the formalism was developed in \cite{gorini2008tolman}, and this can be leveraged to probe the stability of wormholes in terms of the hydrostatic, gravitational, and anisotropic forces in the space-time. The generalized TOV equation \cite{gorini2008tolman,Kuhfittig:2020fue,ponce1993limiting} is given as\\
    
    \begin{equation}\label{eq:tov}
    -\frac{dp_{r}}{dr}-\frac{\epsilon'(r)}{2}(\rho+p_{r})+\frac{2}{r}(p_{t}-p_{r})=0,
    \end{equation}
    
    where $\epsilon(r)=2\Phi(r)$. $F_h$ represents the hydrostatic force, $F_g$ the gravitational force, and $F_a$ the anisotropic force. These three terms of the TOV equation can determine the equilibrium anisotropic mass distribution \cite{ponce1993limiting} in that stable stellar structures satisfy Eq. \eqref{eq:tov}.\\
    
    \begin{equation}\label{stabcomp}
    F_{\mathrm{h}}=-\frac{dp_{r}}{dr},\;\;\;\;\;\;\;\;F_{\mathrm{a}}=\frac{2}{r}(p_{t}-p_{r}), \;\;\;\;\;\;\;\;F_{\mathrm{g}}=-\frac{\epsilon^{'}}{2}(\rho+p_{r})
    \end{equation}
    Figure \ref{fig:seventeen} shows the components of the TOV equation, $F_{\mathrm{h}}$ and $F_{\mathrm{a}}$, for Case I and Case II. It is evident from Eqs. \eqref{eq:tov} and \eqref{stabcomp} that as $\Phi'(r)=0$, the gravitational force $F_{\mathrm{g}}=0$. Figure \ref{fig:seventeen} shows that the hydrostatic force $F_{\mathrm{h}}$ and anisotropic force $F_{\mathrm{a}}$ cancel each other out, rendering a stable wormhole configuration for both Cases I and II.\\
    \vspace{1cm}
    \begin{center}
    \captionof{table}{Summary of the energy conditions}
    \begin{tabular}{ |c|c|c|c| } 
    \hline
    Case & Term & Result & Interpretation \\
    \hline
    \multirow{4}{4em}{I} & \makecell{$\rho$} & \makecell{$>0, \, \forall r$\\} & \makecell{WEC satisfied} \\ 
    & \makecell{$\rho + p_r$} & \makecell{$<0$, \\ near throat} & \makecell{NEC violated at throat} \\
    & \makecell{$\rho+ p_t$} & \makecell{$> 0$, \\ near throat} & \makecell{NEC satisfied at throat} \\
    & \makecell{$\rho + p_r + 2 p_t$} & \makecell{$= 0$, \\ near throat} & \makecell{SEC satisfied at throat} \\
    \hline
    \multirow{4}{4em}{II} & \makecell{$\rho$} & \makecell{$>0, \, \forall r$\\} & \makecell{WEC satisfied} \\ 
    & \makecell{$\rho + p_r$} & \makecell{$<0$, \\ near throat} & \makecell{NEC violated at throat} \\
    & \makecell{$\rho+ p_t$} & \makecell{$> 0$, \\ near throat} & \makecell{NEC satisfied at throat} \\
    & \makecell{$\rho + p_r + 2 p_t$} & \makecell{$< 0$, \\ near throat} & \makecell{SEC violated at throat} \\ 
    \hline
    \multirow{4}{4em}{III} & \makecell{$\rho$} & \makecell{$>0, \, \forall r$, \\ and $\chi = 0.5,1.5,2$} & \makecell{WEC satisfied} \\ 
    & \makecell{$\rho + p_r$} & \makecell{$<0$, \\ near throat for $\chi = 0.5,1.5,2$} & \makecell{NEC violated at throat} \\
    & \makecell{$\rho+ p_t$} & \makecell{$> 0$, \\ near throat for $\chi = 0.5,1.5,2$} & \makecell{NEC satisfied at throat} \\
    & \makecell{$\rho + p_r + 2 p_t$} & \makecell{$< 0$, \\ near throat for $\chi = 0.5,1.5,2$} & \makecell{SEC violated at throat} \\
    \hline
    \multirow{4}{4em}{IV} & \makecell{$\rho$} & \makecell{$>0, \, \forall r$, \\ and $\lambda = -12$} & \makecell{WEC satisfied} \\ 
    & \makecell{$\rho + p_r$} & \makecell{$<0$, \\ near throat for $\lambda = -12$} & \makecell{NEC violated at throat} \\
    & \makecell{$\rho+ p_t$} & \makecell{$> 0$, \\ near throat for $\lambda = -12$} & \makecell{NEC satisfied at throat} \\
    & \makecell{$\rho + p_r + 2 p_t$} & \makecell{$< 0$, \\ near throat for $\lambda = -12$} & \makecell{SEC violated at throat} \\
    \hline
    \end{tabular}
    \label{tab:one}
    \end{center}

\unskip
\begin{figure}[H]
\includegraphics[width=\textwidth]{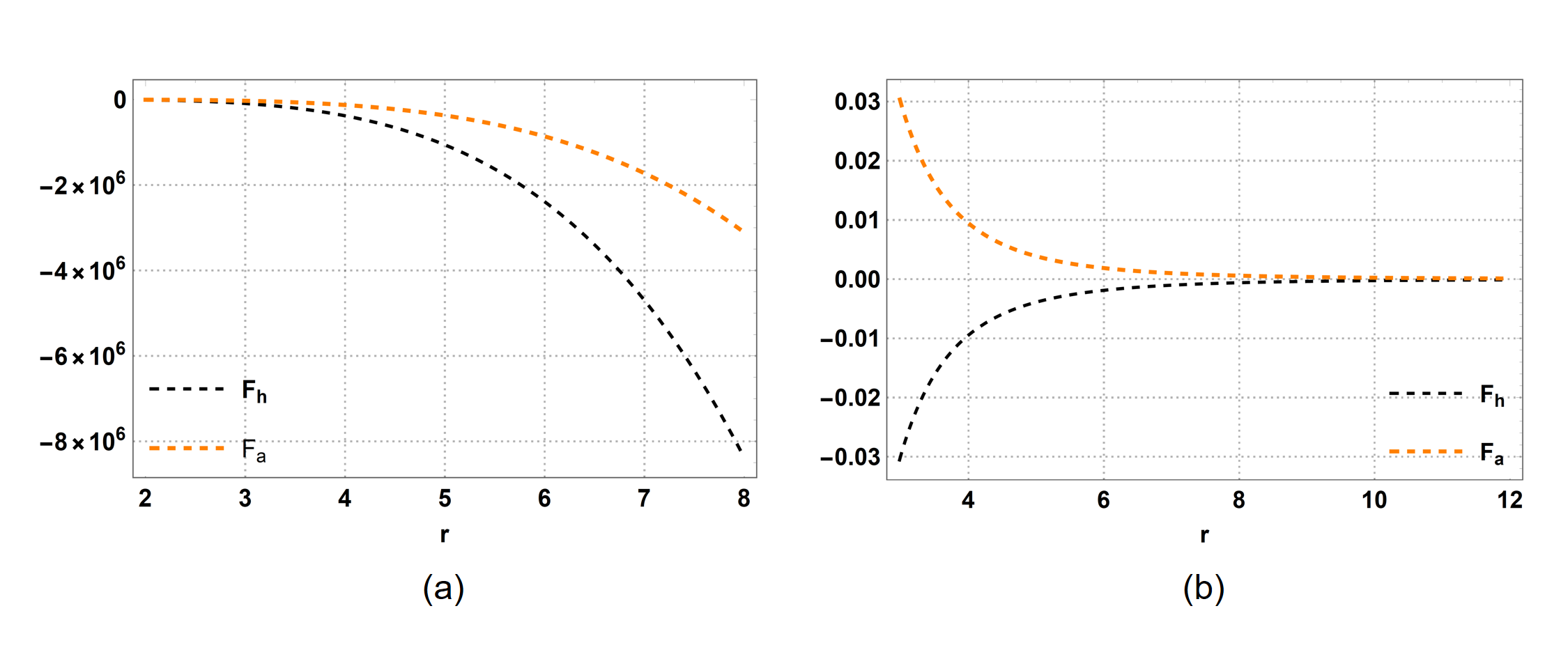}
\caption{Profile of $F_{\mathrm{h}}$ and $F_{\mathrm{a}}$ vs. $r$ with $r_0=0.9$, and $B=0.1$ for (a) Case III with $n=1.1$ and (b) Case IV with $\lambda=-12$.\label{fig:eighteen}}
\end{figure}

    Next, Figure \ref{fig:eighteen} shows the corresponding terms of the TOV equations for Cases III and IV. It can be observed from Figure \ref{fig:eighteen} that the wormhole configuration in Case III ($f(R)$ gravity) is not stable, as the hydrostatic force $F_{\mathrm{h}}$ and anisotropic force $F_{\mathrm{a}}$ do not cancel each other. However, in Case IV ($f(R,T)$ gravity), it can be seen that the anisotropic force $F_{\mathrm{a}}$ and hydrostatic force $F_{\mathrm{h}}$ cancel each other out, thus rendering a stable wormhole configuration.
        \begin{figure}[H]
\includegraphics[width=\textwidth]{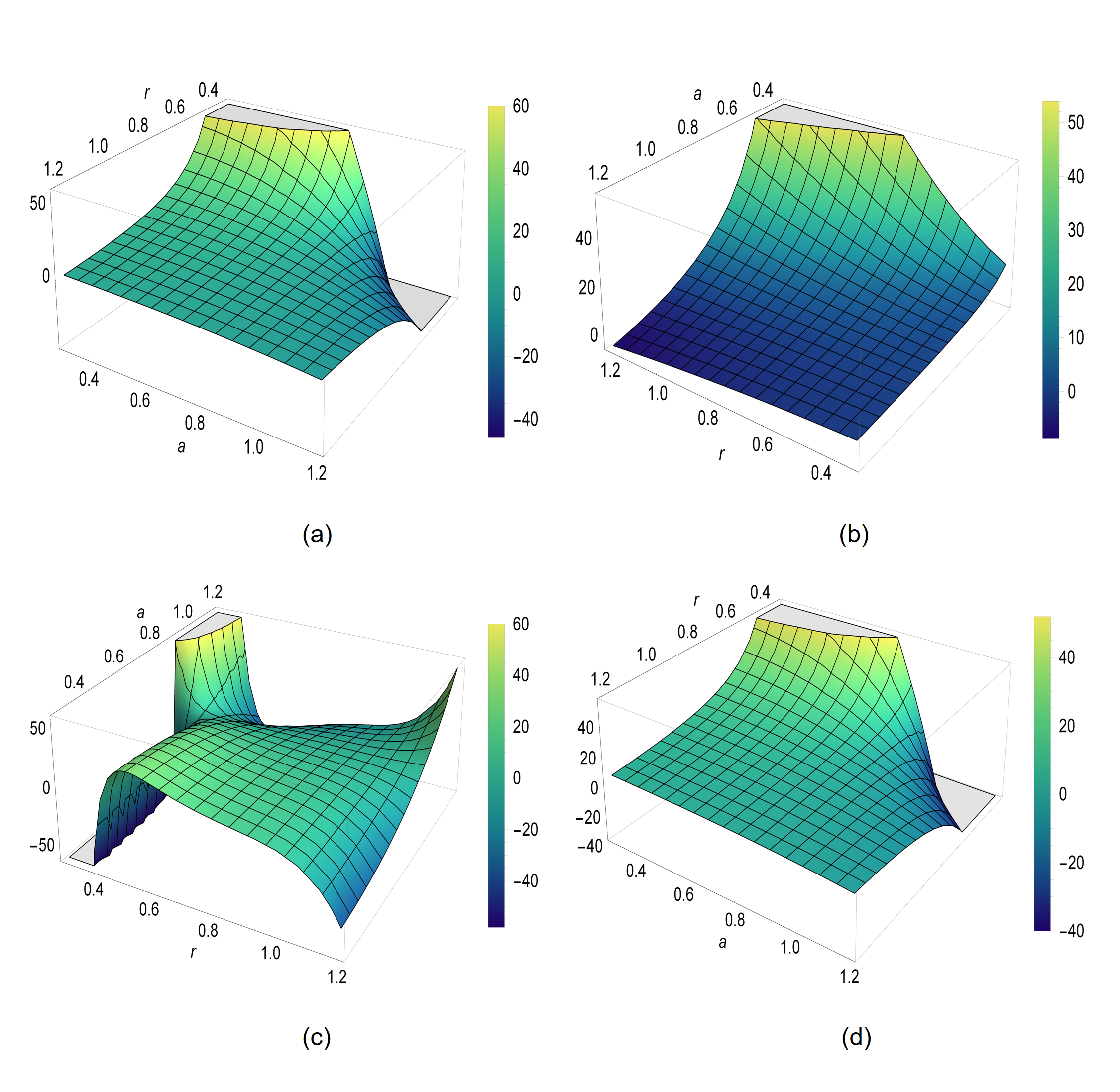}
\caption{Profile of the VIQ for (a) Case I (b) Case II (c) Case III and (d) Case IV.\label{fig:nineteen}}
\end{figure}
\unskip    
    Further, the amount of exotic matter in wormhole space-times can be estimated using the averaged null energy condition, $\int_{\lambda_1}^{\lambda_2} T_{ij}k^ik^j d \lambda \geq 0$, evaluated along the radial coordinate $r$. However, instead of this line integral, a more generalized description of the amount of energy condition violating matter can be estimated by using a volume integral, namely the volume integral quantifier (VIQ) \cite{visser2003traversable,kar2004quantifying,lobo2013new}, which is defined as,
    \begin{equation}
    \label{VIQ}
    I_v = \oint [\rho+p_r] dV = 8\pi\int_{r_0}^{a}(\rho+p_r)r^2 dr
    \end{equation}
    The VIQ gives an estimate of the amount of NEC violating matter required for wormhole configurations after an exterior metric is matched with the wormhole space-time with the stress-energy tensor cutting off at some $r=a$ away from the throat. The requirement of arbitrarily small quantities of NEC violating matter is satisfied provided $I_v \rightarrow 0$ as $a \rightarrow r_0$ \cite{visser2003traversable,kar2004quantifying}. It can be observed from the Figure \ref{fig:nineteen} that as $a \rightarrow r_0$, then, $I_v \rightarrow 0$, signifying that the wormhole configurations may be possible with arbitrarily small amounts of exotic matter for all the four cases.

\section{Discussion}\label{sec4}

In the previous sections, we presented results concerning the energy and stability conditions for wormhole space-times in GR and modified gravity, and it was shown that we obtain characteristic violations of the NEC for all solutions. The SEC is marginally satisfied for Case I (minimally coupled scenario) as expected, but is violated near the throat for the other cases. In the case of non-minimal coupling in GR, we presented a novel approach of coupling, and showed that these results well-approximate the expected behavior for wormhole space-times. A detailed analysis of the energy conditions for all the four cases can be found in Table \ref{tab:one}. It is interesting to note that for the estimated shape functions to satisfy the necessary constraints, the constant`$B$' has to be of the order of $0.01$ for Cases I and II, whereas, for Case III and IV, the constant `$B$' has to be of the order of $0.1$. The constants $B$ were constrained by focusing on the viability of the shape function, since the dimensionality could not be commented on uniquely. Further, from the analyses of the four cases, it is evident that the pseudoscalar axion $H$ as a matter source may yield wormhole configurations in GR as well as in modified gravity. As stated previously, the dual representation of the setup considered here is equivalent to the coupling of a non-self interacting scalar field, and it is worth noting here some remarks concerning wormhole space-times sourced by scalars. Several solutions to Einstein--scalar systems describing wormholes are well-known. For instance, the Ellis--Bronnikov space-time described previously actually describes a symmetric wormhole with a massless scalar. A detailed account on the existence and stability of scalar--sourced wormholes can be found in Refs. \cite{bronnikovmdpi, barandvis}. Butcher \cite{butcherPRD} has shown that one cannot obtain asymptotically well-behaved wormholes in GR that can be supported by scalar fields and non-exotic matter, which is proven for all values of coupling constants. This aspect has also been studied in detail by Barcelo and Visser in Ref. \cite{barandvis}. Here, we have shown that the KR field strength as a source also violates the energy conditions when constrained by the geometric requirements of traversable wormhole space--times. Further, considering the non-minimal coupling serving as a toy model considered here, one can infer that a conformally coupled scenario ($\xi$ = 1/6) can serve as a natural setting for massless modes in curved space--times, as in the invariance of the electromagnetic field. To this end, it is seen that the model considered here in Section \ref{sec:nonmin} also violates the NEC\footnote{We have verified the results with $\xi = 1/6$.}. Further, in Case III, it was observed from the analyses of the TOV equations that the psuedoscalar axion $H$ renders an unstable wormhole configuration in the framework of $f(R)$ gravity, with $f(R) = \chi(n) R^n$. The results in our work depict a clear picture of possible wormhole configurations sourced by the KR field, and should serve as reference points for future numerical or analytical studies in this direction. There are some limitations to the current work. First, the stability of the proposed space-times against perturbations has not been discussed, and requires further analytical investigation. Additionally, the power-law $f(R)$ gravity model considered in this work is one of the simpler corrections to GR, and more feasible models can be investigated to probe the issue of stable/unstable KR field-sourced wormholes in modified gravity. Further, a possible future direction is the estimation of potentially observable signatures such as quasinormal modes, shadows, and evolution of null-geodesics from these space-times, and further constraining the parameters using stringent observational bounds. In the context of Einstein-KR gravity, static and non-static spherically symmetric black holes have been reported in both the minimal and non-minimally coupled scenario, and properties such as thermodynamics and evolution of null geodesics have been analyzed. To this end, it is an interesting question to distinguish such properties between black hole and wormhole-like solutions, especially considering the possibility of black hole-mimicking exotic ultra-compact objects such as wormholes. These are potential issues of interest in the context of KR gravity to be reported in the near-future.
\\
Although wormholes have not been observed till date, these solutions have significant implications in several problems in gravity such as the cosmic censorship hypothesis \cite{penrose2002golden}, $ER=EPR$ paradigm \cite{epr}, paradoxes involving closed time-like curves etc. Gravity theories with quantum approximations also yield wormhole-like solutions, and serve as a ground for testing deviations from GR in the near future with advances in multi-messenger astronomy. Thus, further progress in wormhole studies and detection techniques is crucial for understanding the fundamental nature of the universe and space-time.
   \subsection*{Acknowledgement}
   The authors would like to extend their gratitude to Dr. Sayantan Choudhury, The Thanu Padmanabhan Centre for Cosmology and Science Popularization, for helpful discussions regarding KR field theory.
    \section*{References}
    \printbibliography[heading=none]

@article{hehl1976general,
	title        = {General relativity with spin and torsion: Foundations and prospects},
	author       = {Hehl, F. W. and V. der H., Paul and Kerlick, G. D. and Nester, J. M.},
	year         = 1976,
	journal      = {Reviews of Modern Physics},
	publisher    = {APS},
	volume       = 48,
	number       = 3,
	pages        = 393
}

@article{raychaudhuri1979theoretical,
	title        = {Theoretical cosmology},
	author       = {Raychaudhuri, A. K.},
	year         = 1979,
	journal      = {Oxford Studies in Physics}
}

@article{kalb1974classical,
	title        = {Classical direct interstring action},
	author       = {Kalb, M. and Ramond, P.},
	year         = 1974,
	journal      = {Physical Review D},
	publisher    = {APS},
	volume       = 9,
	number       = 8,
	pages        = 2273
}

@article{sengupta2001spherically,
	title        = {Spherically symmetric solutions of gravitational field equations in Kalb--Ramond background},
	author       = {SenGupta, S. and Sur, S.},
	year         = 2001,
	journal      = {Physics Letters B},
	publisher    = {Elsevier},
	volume       = 521,
	number       = {3-4},
	pages        = {350--356}
}

@article{kar2003static,
	title        = {Static, spherically symmetric solutions, gravitational lensing, and perihelion precession in Einstein-Kalb-Ramond theory},
	author       = {Kar, S. and SenGupta, S. and Sur, S.},
	year         = 2003,
	journal      = {Physical Review D},
	publisher    = {APS},
	volume       = 67,
	number       = 4,
	pages        = {044005}
}

@article{baruah2023non,
	title        = {Non-commutative wormholes in f (R) gravity satisfying the energy conditions},
	author       = {Baruah, A. and Goswami, P. and Deshamukhya, A.},
	year         = 2023,
	journal      = {New Astronomy},
	publisher    = {Elsevier},
	volume       = 99,
	pages        = 101956
}

@article{baruah2022new,
	title        = {New wormhole solutions in a viable f (R) gravity model},
	author       = {Baruah, A. and Goswami, P. and Deshamukhya, A.},
	year         = 2022,
	journal      = {International Journal of Modern Physics D},
	publisher    = {World Scientific},
	pages        = 2250119
}

@article{harko2011f,
	title        = {f (R, T) gravity},
	author       = {Harko, T. and Lobo, F. S. N. and Nojiri, S. and Odintsov, S. D.},
	year         = 2011,
	journal      = {Physical Review D},
	publisher    = {APS},
	volume       = 84,
	number       = 2,
	pages        = {024020}
}

@article{chakraborty2018packing,
	title        = {Packing extra mass in compact stellar structures: An interplay between Kalb-Ramond field and extra dimensions},
	author       = {Chakraborty, S. and SenGupta, S.},
	year         = 2018,
	journal      = {Journal of Cosmology and Astroparticle Physics},
	publisher    = {IOP Publishing},
	volume       = 2018,
	number       = {05},
	pages        = {032}
}

@article{cox2016stability,
	title        = {Stability of Einstein-Maxwell-Kalb-Ramond wormholes},
	author       = {Cox, P. H. and Harms, B. C. and Hou, S.},
	year         = 2016,
	journal      = {Physical Review D},
	publisher    = {APS},
	volume       = 93,
	number       = 4,
	pages        = {044014}
}

@article{maity2004parity,
	title        = {Parity-violating Kalb--Ramond--Maxwell interactions and CMB anisotropy in a braneworld},
	author       = {Maity, D. and Majumdar, P. and SenGupta, S.},
	year         = 2004,
	journal      = {Journal of Cosmology and Astroparticle Physics},
	publisher    = {IOP Publishing},
	volume       = 2004,
	number       = {06},
	pages        = {005}
}

@article{carloni2005cosmological,
	title        = {Cosmological dynamics of $R^n$ gravity},
	author       = {Carloni, S. and Dunsby, P. K. S. and Capozziello, S. and Troisi, A.},
	year         = 2005,
	journal      = {Classical and Quantum Gravity},
	publisher    = {IOP Publishing},
	volume       = 22,
	number       = 22,
	pages        = 4839
}

@article{nojiri2007introduction,
	title        = {Introduction to modified gravity and gravitational alternative for dark energy},
	author       = {Nojiri, S. and Odintsov, S. D.},
	year         = 2007,
	journal      = {International Journal of Geometric Methods in Modern Physics},
	publisher    = {World Scientific},
	volume       = 4,
	number       = {01},
	pages        = {115--145}
}

@article{albareti2013non,
	title        = {On the non-attractive character of gravity in f (R) theories},
	author       = {Albareti, F. D. and Cembranos, J. A. R. and De La Cruz-Dombriz, A. and Dobado, A.},
	year         = 2013,
	journal      = {Journal of Cosmology and Astroparticle Physics},
	publisher    = {IOP Publishing},
	volume       = 2013,
	number       = {07},
	pages        = {009}
}

@book{Visser:1995cc,
	title        = {{Lorentzian wormholes: From Einstein to Hawking}},
	author       = {Visser, M.},
	year         = 1995,
	publisher    = {Woodbury},
	address      = {{USA}}
}

@article{Morris:1988cz,
	title        = {{Wormholes in space-time and their use for interstellar travel: A tool for teaching general relativity}},
	author       = {Morris, M. S. and Thorne, K. S.},
	year         = 1988,
	journal      = {American Journal of Physics},
	volume       = 56,
	pages        = {395--412}
}

@article{PhysRevD.80.104012,
	title        = {Wormhole geometries in {$f(R)$} modified theories of gravity},
	author       = {F. S. N. Lobo and M. A. Oliveira},
	year         = 2009,
	journal      = {Physical Review D},
	publisher    = {American Physical Society},
	volume       = 80,
	pages        = 104012,
	issue        = 10,
	numpages     = 9
}

@article{pavlovic2015wormholes,
	title        = {Wormholes in viable {$f(R)$} modified theories of gravity and weak energy condition},
	author       = {P. Pavlovic and M. Sossich},
	year         = 2015,
	journal      = {European Physical Journal C},
	publisher    = {Springer},
	volume       = 75,
	number       = 3,
	pages        = {1--8}
}

@article{PhysRevD.96.044038,
	title        = {Modeling wormholes in {$f(R,T)$} gravity},
	author       = {P. Moraes and P. K. Sahoo},
	year         = 2017,
	journal      = {Physical Review D},
	publisher    = {American Physical Society},
	volume       = 96,
	pages        = {044038},
	issue        = 4,
	numpages     = 8
}

@article{PhysRevD.94.044041,
	title        = {Cosmological wormholes in {$f(R)$} theories of gravity},
	author       = {S. Bahamonde and M. Jamil and P. Pavlovic and M. Sossich},
	year         = 2016,
	journal      = {Physical Review D},
	publisher    = {American Physical Society},
	volume       = 94,
	pages        = {044041},
	issue        = 4,
	numpages     = 12
}

@article{doi:10.1142/S0218271820500686,
	title        = {Traversable wormholes with exponential shape function in modified gravity and general relativity: A comparative study},
	author       = {G. C. Samanta and N. Godani and K. Bamba},
	year         = 2020,
	journal      = {International Journal of Modern Physics D},
	volume       = 29,
	number       = {09},
	pages        = 2050068
}

@article{azizi2013wormhole,
	title        = {Wormhole geometries in {$f(R,T)$} gravity},
	author       = {T. Azizi},
	year         = 2013,
	journal      = {International Journal of Theoretical Physics},
	publisher    = {Springer},
	volume       = 52,
	number       = 10,
	pages        = {3486--3493}
}

@incollection{curiel2017primer,
	title        = {A primer on energy conditions},
	author       = {E. Curiel},
	year         = 2017,
	booktitle    = {Towards a theory of spacetime theories},
	publisher    = {Springer New York},
	address      = {New York, NY},
	pages        = {43--104}
}

@article{penrose2002golden,
	title        = {“{Golden Oldie}”: Gravitational collapse: the role of general relativity},
	author       = {R. Penrose},
	year         = 2002,
	journal      = {General Relativity and Gravitation},
	publisher    = {Citeseer},
	volume       = 34,
	number       = 7,
	pages        = {1141--1165}
}

@article{epr,
	title        = {{ER=EPR}, {GHZ}, and the consistency of quantum measurements},
	author       = {L. Susskind},
	year         = 2016,
	journal      = {Fortschritte der Physik},
	volume       = 64,
	number       = 1,
	pages        = {72--83}
}

@article{oppenheimer1939massive,
	title        = {On massive neutron cores},
	author       = {R. J. Oppenheimer and G. M. Volkoff},
	year         = 1939,
	journal      = {Physical Review},
	volume       = 55,
	number       = 4,
	pages        = 374
}

@article{gorini2008tolman,
	title        = {Tolman-Oppenheimer-Volkoff equations in the presence of the Chaplygin gas: Stars and wormholelike solutions},
	author       = {Gorini, V. and Moschella, U. and Kamenshchik, A. Yu and Pasquier, V. and Starobinsky, A. A.},
	year         = 2008,
	journal      = {Physical Review D},
	volume       = 78,
	number       = 6,
	pages        = {064064}
}

@article{Kuhfittig:2020fue,
	title        = {A note on the stability of Morris-Thorne wormholes},
	author       = {Kuhfittig, P. K. F.},
	year         = 2020,
	journal      = {Fundamental Journal of Modern Physics},
	volume       = 14,
	pages        = {23--31}
}

@article{ponce1993limiting,
	title        = {Limiting configurations allowed by the energy conditions},
	author       = {J. Ponce de Leon},
	year         = 1993,
	journal      = {General relativity and gravitation},
	volume       = 25,
	number       = 11,
	pages        = {1123--1137}
}

@article{visser2003traversable,
	title        = {Traversable wormholes with arbitrarily small energy condition violations},
	author       = {M. Visser and S. Kar and N. Dadhich},
	year         = 2003,
	journal      = {Physical Review letters},
	volume       = 90,
	number       = 20,
	pages        = 201102
}

@article{kar2004quantifying,
	title        = {Quantifying energy condition violations in traversable wormholes},
	author       = {S. Kar and N. Dadhich and M. Visser},
	year         = 2004,
	journal      = {Pramana},
	volume       = 63,
	number       = 4,
	pages        = {859--864}
}

@article{lobo2013new,
	title        = {New asymptotically flat phantom wormhole solutions},
	author       = {F. S. N. Lobo and F. Parsaei and N. Riazi},
	year         = 2013,
	journal      = {Physical Review D},
	volume       = 87,
	number       = 8,
	pages        = {084030}
}

@article{sankharva2022nonminimally,
	title        = {Nonminimally coupled ultralight axions as cold dark matter},
	author       = {Sankharva, K. and Sethi, S.},
	year         = 2022,
	journal      = {Physical Review D},
	publisher    = {APS},
	volume       = 105,
	number       = 10,
	pages        = 103517
}

@article{harko2014thermodynamic,
	title        = {Thermodynamic interpretation of the generalized gravity models with geometry-matter coupling},
	author       = {Harko, T.},
	year         = 2014,
	journal      = {Physical Review D},
	publisher    = {APS},
	volume       = 90,
	number       = 4,
	pages        = {044067}
}

@inproceedings{cartan1923varietes,
	title        = {Sur les vari{\'e}t{\'e}s {\`a} connexion affine et la th{\'e}orie de la relativit{\'e} g{\'e}n{\'e}ralis{\'e}e (premi{\`e}re partie)},
	author       = {Cartan, {\'E}.},
	year         = 1923,
	booktitle    = {Annales scientifiques de l'{\'E}cole normale sup{\'e}rieure},
	volume       = 40,
	pages        = {325--412}
}

@inproceedings{cartan1924varietes,
	title        = {Sur les vari{\'e}t{\'e}s {\`a} connexion affine, et la th{\'e}orie de la relativit{\'e} g{\'e}n{\'e}ralis{\'e}e (premi{\`e}re partie)(suite)},
	author       = {Cartan, {\'E}.},
	year         = 1924,
	booktitle    = {Annales scientifiques de l'{\'E}cole Normale Sup{\'e}rieure},
	volume       = 41,
	pages        = {1--25}
}

@inproceedings{cartan1925varietes,
	title        = {Sur les vari{\'e}t{\'e}s {\`a} connexion affine, et la th{\'e}orie de la relativit{\'e} g{\'e}n{\'e}ralis{\'e}e (deuxi{\`e}me partie)},
	author       = {Cartan, {\'E}.},
	year         = 1925,
	booktitle    = {Annales scientifiques de l'{\'E}cole normale sup{\'e}rieure},
	volume       = 42,
	pages        = {17--88}
}

@book{cartan1986manifolds,
	title        = {{On Manifolds with an Affine Connection and the Theory of General Relativity, translated by A. Magon and A. Ashtekar}},
	author       = {Cartan, {\'E}.},
	year         = 1986,
	publisher    = {Bibiliopolis},
	address      = {{Napoli, Italy}}
}

@article{majumdar1999parity,
	title        = {Parity-violating gravitational coupling of electromagnetic fields},
	author       = {Majumdar, P. and SenGupta, S.},
	year         = 1999,
	journal      = {Classical and Quantum Gravity},
	publisher    = {IOP Publishing},
	volume       = 16,
	number       = 12,
	pages        = {L89}
}

@article{shabani2020connection,
	title        = {A connection between Rastall-type and f (R, T) gravities},
	author       = {Shabani, H. and Ziaie, A. H.},
	year         = 2020,
	journal      = {Europhysics Letters: EPL},
	publisher    = {IOP Publishing},
	volume       = 129,
	number       = 2,
	pages        = 20004
}

@article{Baruah_2019,
	title        = {Traversable Lorentzian wormholes in higher dimensional theories of gravity},
	author       = {A. Baruah and A. Deshamukhya},
	year         = 2019,
	month        = {oct},
	journal      = {Journal of Physics: Conference Series},
	publisher    = {{IOP} Publishing},
	volume       = 1330,
	number       = 1,
	pages        = {012001}
}

@article{izmailov2019can,
	title        = {Can massless wormholes mimic a Schwarzschild black hole in the strong field lensing?},
	author       = {Izmailov, R. N. and Bhattacharya, A. and Zhdanov, E. R. and Potapov, A. A. and Nandi, K. K.},
	year         = 2019,
	journal      = {European Physical Journal Plus},
	publisher    = {Springer Berlin Heidelberg},
	volume       = 134,
	number       = 8,
	pages        = 384
}

@article{nandi2017ring,
	title        = {Ring-down gravitational waves and lensing observables: How far can a wormhole mimic those of a black hole?},
	author       = {Nandi, K. K and Izmailov, R. N and Yanbekov, A. A. and Shayakhmetov, A. A.},
	year         = 2017,
	journal      = {Physical Review D},
	publisher    = {APS},
	volume       = 95,
	number       = 10,
	pages        = 104011
}

@article{konoplya2022traversable,
	title        = {Traversable Wormholes in General Relativity},
	author       = {Konoplya, R. A. and Zhidenko, A.},
	year         = 2022,
	journal      = {Physical Review Letters},
	publisher    = {APS},
	volume       = 128,
	number       = 9,
	pages        = {091104}
}

@article{karakasis2022f,
	title        = {$f(R)$ gravity wormholes sourced by a phantom scalar field},
	author       = {Karakasis, T. and Papantonopoulos, E. and Vlachos, C.},
	year         = 2022,
	journal      = {Physical Review D},
	publisher    = {APS},
	volume       = 105,
	number       = 2,
	pages        = {024006}
}

@article{kostelecky1989spontaneous,
	title        = {Spontaneous breaking of Lorentz symmetry in string theory},
	author       = {Kosteleck{\`y}, V. A. and Samuel, S.},
	year         = 1989,
	journal      = {Physical Review D},
	publisher    = {APS},
	volume       = 39,
	number       = 2,
	pages        = 683
}

@article{altschul2010lorentz,
	title        = {Lorentz violation with an antisymmetric tensor},
	author       = {Altschul, B. and Bailey, Q. G. and Kosteleck{\`y}, V. A.},
	year         = 2010,
	journal      = {Physical Review D},
	publisher    = {APS},
	volume       = 81,
	number       = 6,
	pages        = {065028}
}

@article{lessa2021traversable,
	title        = {Traversable wormhole solution with a background Kalb–Ramond field},
	author       = {Lessa, L. A. and Oliveira, R and Silva, J. E. G. and Almeida, C. A. S.},
	year         = 2021,
	journal      = {Annals of Physics},
	publisher    = {Elsevier},
	volume       = 433,
	pages        = 168604
}

@article{clifton2005power,
	title        = {The power of general relativity},
	author       = {Clifton, T. and Barrow, J. D.},
	year         = 2005,
	journal      = {Physical Review D},
	publisher    = {APS},
	volume       = 72,
	number       = 10,
	pages        = 103005
}

@article{sokoliuk2022probing,
	title        = {Probing the existence of the ZTF Casimir wormholes in the framework of $f(R)$ gravity},
	author       = {Sokoliuk, O. and Baransky, A. and Sahoo, P. K.},
	year         = 2022,
	journal      = {Nuclear Physics B},
	publisher    = {Elsevier},
	pages        = 115845
}

@article{ellis1973ether,
  title={Ether flow through a drainhole: A particle model in general relativity},
  author={Ellis, H. G.},
  journal={Journal of Mathematical Physics},
  volume={14},
  number={1},
  pages={104--118},
  year={1973},
  publisher={American Institute of Physics}
}

@article{bronnikov1973scalar,
  title={Scalar-tensor theory and scalar charge},
  author={Bronnikov, K. A.},
  journal={Acta Phys. Pol},
  pages={B4},
  year={1973}
}

@article{butcherPRD,
  title={Traversable wormholes and classical scalar fields},
  author={Butcher, L. M.},
  journal={Physical Review D},
  volume={91},
  number={12},
  pages={124031},
  year={2015},
  publisher={APS}
}

@article{barandvis,
  title={Scalar fields, energy conditions and traversable wormholes},
  author={Barcelo, C. and Visser, M.},
  journal={Classical and Quantum Gravity},
  volume={17},
  number={18},
  pages={3843},
  year={2000},
  publisher={IOP Publishing}
}

@article{bronnikovmdpi,
  title={Scalar fields as sources for wormholes and regular black holes},
  author={Bronnikov, K. A.},
  journal={Particles},
  volume={1},
  number={1},
  pages={56--81},
  year={2018},
  publisher={MDPI}
}

@book{tolman1987relativity,
  title={Relativity, thermodynamics, and cosmology},
  author={Tolman, R. C.},
  year={1987},
  publisher={Courier Corporation}
}

\end{document}